\documentclass{elsartL}
\usepackage{graphicx}
\usepackage{amssymb}
\usepackage{amsmath}
\usepackage{mathrsfs}
\usepackage{mathtools}
\newcommand{\avg}[1]{\left< #1 \right>} 
\begin{document}

\begin{frontmatter}
\title{What has been learnt from the analysis of the low-energy pion-nucleon data during the past three decades?}
\author{Evangelos Matsinos}
\address{Physik-Institut, Universit{\"a}t Z{\"u}rich, CH-8057 Z{\"u}rich, Switzerland}
\begin{abstract}
Over twenty-five years ago, two analyses of the pion-nucleon ($\pi N$) data at low energy (i.e., for pion laboratory kinetic energy $T \leq 100$ MeV) reported on the departure of the extracted scattering amplitudes, 
corresponding to the two elastic-scattering reactions $\pi^\pm p \to \pi^\pm p$ and to the $\pi^- p$ charge-exchange reaction $\pi^- p \to \pi^0 n$, from the triangle identity, which these amplitudes fulfil if the isospin 
invariance holds in the hadronic part of the $\pi N$ interaction. This discrepancy indicates that \emph{at least one} of the following assumptions is not valid: first, that the absolute normalisation of the bulk of the 
low-energy $\pi N$ datasets is correct; second, that any residual contributions to the corrections, which aim at the removal of the effects of electromagnetic (EM) origin from the measurements, are not significant; and third, 
that the isospin invariance holds in the hadronic part of the $\pi N$ interaction. In view of the incompatibility of the results of the various schemes of removal of the so-called trivial EM effects at the $\pi N$ threshold 
($T = 0$ MeV), the likelihood of residual effects of EM origin in the extracted $\pi N$ scattering amplitudes (from the data in the scattering region, i.e., above the $\pi N$ threshold) must be reassessed. This work 
emphasises the importance of the development of a unified scheme for the determination of reliable EM corrections, \emph{applicable at the $\pi N$ threshold and in the scattering region}, in providing a resolution to the 
established discrepancy at low energy.
\end{abstract}
\end{frontmatter}

\section{\label{sec:Introduction}Introduction}

It was about mid November 2021 when Igor Strakovsky came up with the idea of a new book, summarising our knowledge of the properties of the pion, and kindly asked me to relate in that book my experience from the analysis 
of the pion-nucleon ($\pi N$) experimental data at low energy. I found his idea appealing and, after a day or two, I made my decision to join the `celebration of the $75$ years since the discovery of the pion', and to 
attempt to explain to a broad, or - more accurately - to a broader than usual, audience of physicists why the interaction between pions and nucleons is interesting at all.

I ought to admit that my initial decision to take part in this effort was followed by a short interval of vacillation regarding the optimal organisation and presentation of the material. The trade-off between what I (as an 
author) expected of the average reader of Igor's book versus what the average reader expected of me urged me to abandon mathematical rigour, to depart from my `obsession' with technicalities, and to attempt to recount my 
experience in a more engaging manner, while retaining both accuracy and simplicity. Before commencing however, I recommend Ref.~\cite{Matsinos2017a}, as well as the works cited therein, to those of the readers who are keen 
on the mathematical details of this research programme, as well as on its development over time. As I started this project when I was a postdoctoral researcher at the Swiss Federal Institute of Technology in Zurich 
(Eidgen{\"o}ssische Technische Hochschule Z{\"u}rich), I will refer to it as `the ETH $\pi N$ project' henceforth.

I recently decided to make this work publicly available as a preprint. Unlikely as it might seem, my hope remains that the broader dissemination of this report (as opposed to its restricted function as a chapter in a book) 
will rekindle interest in the Pion-Physics domain.

Overall, Physics researchers are hardly amongst the most talented when it comes down to communicating efficiently the crux of their studies, even to colleagues as the case frequently is. In an effort to break this pattern, 
I will present the material in the form of a series of questions and answers in parts of this work; I have been asked these questions many times. This `didactical' distraction averts a tedious monologue and brings the reader 
in plain view as an interlocutor.

The structure of this paper is as follows. After I discuss what makes the analysis of the $\pi N$ measurements at low energy interesting in Section \ref{sec:Preliminaries}, I provide some details about the experimental data. 
In Section \ref{sec:Analysis}, I describe the steps, which lead from the experimental data to the extraction of the hadronic part of the $\pi N$ interaction: these steps involve the modelling of that part, the inclusion of 
the electromagnetic (EM) effects, the selection of the input database (DB), and the selection of the optimisation procedure. In the subsequent section, I give a summary of what I have learnt from the various analyses of the 
low-energy $\pi N$ data thus far; that section establishes a disturbing discrepancy, called herein the `low-energy $\pi N$ enigma'. Section \ref{sec:Explanations} is bound to be speculative, examining a few straightforward 
possibilities which could explain the observed discrepancies. The conclusions of this work, including a short discussion about the steps towards a resolution to the low-energy $\pi N$ enigma, are set forth in the last section, 
Section \ref{sec:Conclusions}.

In this work, all rest masses of particles and all $3$-momenta will be expressed in energy units. All scattering lengths will be expressed in fm. Whenever two uncertainties accompany a physical result, the first one is 
statistical and the second systematic. The values of the physical constants have been fixed from the 2020 compilation of the Particle-Data Group \cite{PDG2020}.

\section{\label{sec:Preliminaries}Preliminaries}

\subsection{\label{sec:LowEnergy}Why low energy?}

Within the ETH $\pi N$ project, `low energy' implies the restriction of the pion laboratory kinetic energy $T$ of the input experimental data to values up to $100$ MeV. After 1994, all phase-shift analyses (PSAs) in this 
programme have been performed using data from $T=0$ (known as `$\pi N$ threshold', though technically one means the `$\pi^\pm p$ threshold') to $100$ MeV. A few words about this choice are in order. In retrospect, there are 
five reasons why the analyses are limited to the aforementioned energy region.
\begin{itemize}
\item Owing to the experiments conducted at the meson factories over three decades, the low-energy $\pi N$ DB became extensive enough to enable exclusive analyses.
\item In its current form, the hadronic model of this programme (which will be referred to as `ETH model' henceforth) is suitable for PSAs at low $4$-momentum transfer $Q^2$. Although the description of the data even above 
the $\Delta(1232)$ resonance had been (successfully) attempted in the early 1990s, the sensitivity of the results (i.e., of the fitted values of the model parameters) to the inclusion (in the Feynman diagrams of the model) 
of hadronic form factors must be carefully addressed prior to the enhancement of the DB towards higher energy. At present, it can only be asserted that the impact of such effects on the results, obtained from the low-energy 
$\pi N$ DB, is insignificant.
\item It is debatable whether the theoretical constraints, which are valid in the region of asymptotic freedom, also hold at low energy. To estimate the dispersion integrals, all dispersion-relation analyses rely (by and 
large) on high-energy data. The description of the low-energy data in an unbiased manner (i.e., without any high-energy influences) is not possible in such schemes. The exclusive analysis of the low-energy $\pi N$ DB, in 
terms of consistency and compliance with theoretical constraints, such as the isospin invariance in the hadronic part of the $\pi N$ interaction, is therefore interesting in its own sake.
\item Within the ETH $\pi N$ project, the removal of the effects of EM origin from the extracted $\pi N$ phase shifts and partial-wave amplitudes rests upon the application of EM corrections, which are not available for 
$T > 100$ MeV (at least at present).
\item Interest in the low-energy $\pi N$ interaction was maintained for several decades by the prospect of obtaining estimates for the $\pi N$ $\sigma$ term using low-energy input exclusively (e.g., the $\pi N$ phase shifts 
at low energy). It can be argued that the extrapolation of the $\pi N$ scattering amplitudes into the unphysical region is more reliable when it rests upon `close-by' data, thus avoiding any high-energy influences. This 
issue was first addressed (and realised) in Ref.~\cite{Alarcon2012}.
\end{itemize}

\subsection{\label{sec:Observables}Measurable quantities}

As various definitions of the term `dataset' have been in use, involving different choices of the experimental conditions which ought to remain stable/identical during the data acquisition, I feel that, before entering the 
description of the low-energy $\pi N$ DB, I better explain what the term implies in this work. The properties of the incident beam and the (physical, geometrical) characteristics of the target were employed in the past in 
order to distinguish the results of experiments conducted at one place over a (short) period of time. However, datasets have appeared in experimental reports relevant to the $\pi N$ interaction, which not only involved 
different beam energies, but also contained measurements of different reactions. In this work, the requisite for accepting a set of observations as comprising one dataset is that these observations share the same measurement 
of the absolute normalisation~\footnote{Of course, this is a necessary, not a sufficient, condition. Additional requirements may apply after the examination of the original experimental reports, in particular regarding the 
off-line processing of the raw experimental data.} (and, consequently, identical normalisation uncertainty). All references to the experimental works may be found in Ref.~\cite{Matsinos2017a} and in the works cited therein. 
Unless attention must be drawn to a specific dataset, no explicit reference will be given to the original experimental reports.

The composition of the low-energy $\pi N$ DB in terms of number of entries (datapoints), arranged in datasets (following the definition of the previous paragraph), for the various low-energy $\pi N$ measurable quantities 
(observables) is displayed in Table \ref{tab:DB}. The low-energy $\pi N$ DB at finite (non-zero) $T$ comprises measurements of the differential cross section (DCS), analysing power (AP), partial-total cross section (PTCS), 
and total-nuclear cross section (TNCS) for the two elastic-scattering (ES) reactions ($\pi^\pm p \to \pi^\pm p$). One dataset of AP measurements contains data of both ES reactions, namely seven $\pi^+ p$ datapoints and 
three $\pi^- p$ ES datapoints. In addition to the DCS and AP measurements, the DB of the $\pi^- p$ charge-exchange (CX) reaction ($\pi^- p \to \pi^0 n$) contains measurements of the total cross section (TCS). Furthermore, 
two experiments (conducted in the 1980s) measured the $\pi^- p$ CX DCS, but the experimental groups published the corresponding (fitted) values of the first three coefficients in the Legendre expansion (CLE) of their DCSs. 
At present, the two ES reactions and the $\pi^- p$ CX reaction are the only $\pi N$ processes which are experimentally accessible at low energy.

\vspace{0.5cm}
\begin{table}[h!]
{\bf \caption{\label{tab:DB}}}The breakdown of the low-energy $\pi N$ DB into reactions and measured physical quantities. The entries represent the numbers of the datapoints and of the corresponding datasets in the DB. The 
data of this table have appeared in peer-reviewed Physics journals, in the sixteen issues of the $\pi N$ Newsletter, or in dissertations. Not included in this table (but contained in the SAID $\pi N$ DB) are $16$ measurements 
(three datasets) of the $\pi^+ p$ DCS and $22$ measurements (four datasets) of the $\pi^- p$ ES DCS, which had appeared in preprints, which were not submitted/accepted for publication, or had been privately communicated to 
the SAID group.
\vspace{0.25cm}
\begin{center}
\begin{tabular}{|l|c|c|c|c|c|c|c|c|c|}
\hline
\multicolumn{10}{|c|}{Datapoints}\\
\hline
Reaction & DCS & AP & PTCS & TNCS & TCS & CLE & $\epsilon_{1s}$ & $\Gamma_{1s}$ & Total\\
\hline
$\pi^+ p$ & $682$ & $31$ & $24$ & $6$ & $-$ & $-$ & $-$ & $-$ & $743$\\
$\pi^- p$ ES & $525$ & $85$ & $3$ & $6$ & $-$ & $-$ & $2$ & $-$ & $621$\\
$\pi^- p$ CX & $297$ & $10$ & $-$ & $-$ & $10$ & $18$ & $-$ & $2$ & $337$\\
$\pi^+ p$ and $\pi^- p$ ES & $-$ & $10$ & $-$ & $-$ & $-$ & $-$ & $-$ & $-$ & $10$\\
\hline
Total & $1504$ & $136$ & $27$ & $12$ & $10$ & $18$ & $2$ & $2$ & $1711$\\
\hline
\hline
\multicolumn{10}{|c|}{Datasets}\\
\hline
Reaction & DCS & AP & PTCS & TNCS & TCS & CLE & $\epsilon_{1s}$ & $\Gamma_{1s}$ & Total\\
\hline
$\pi^+ p$ & $53$ & $5$ & $19$ & $6$ & $-$ & $-$ & $-$ & $-$ & $83$\\
$\pi^- p$ ES & $40$ & $9$ & $3$ & $6$ & $-$ & $-$ & $2$ & $-$ & $60$\\
$\pi^- p$ CX & $36$ & $2$ & $-$ & $-$ & $10$ & $6$ & $-$ & $2$ & $56$\\
$\pi^+ p$ and $\pi^- p$ ES & $-$ & $1$ & $-$ & $-$ & $-$ & $-$ & $-$ & $-$ & $1$\\
\hline
Total & $129$ & $17$ & $22$ & $12$ & $10$ & $6$ & $2$ & $2$ & $200$\\
\hline
\end{tabular}
\end{center}
\vspace{0.5cm}
\end{table}

Unlike all other analyses of the $\pi N$ DB (which I am aware of), the studies performed within the ETH $\pi N$ project have made use (after the mid 1990s) of the $\pi^+ p$ PTCSs and TNCSs, as well as of the $\pi^- p$ CX 
TCSs. It would have been controversial to include in the DB the nine measurements of the $\pi^- p$ ES PTCSs and TNCSs (see Ref.~\cite{Matsinos2017a} for a discussion).

The $\pi N$ scattering amplitude $\mathscr{F} (\vec{q}^{\, \prime}, \vec{q})$ at low energy may safely be confined to $s$- and $p$-wave contributions (e.g., see Ref.~\cite{Ericson1988}, pp.~17--18). Introducing the isospin 
of the pion as $\vec{t}$ and that of the nucleon as $\vec{\tau}/2$, one may write (using natural units for the sake of brevity):
\begin{equation} \label{eq:EQ001}
\mathscr{F} (\vec{q}^{\, \prime}, \vec{q}) = b_0 + b_1 \, \vec{\tau} \cdot \vec{t} + \left( c_0 + c_1 \, \vec{\tau} \cdot \vec{t} \, \right) \vec{q}^{\, \prime} \cdot \vec{q} + i \left( d_0 + d_1 \, \vec{\tau} \cdot \vec{t} \, \right) \vec\sigma \cdot 
(\vec{q}^{\, \prime} \times \vec{q}) \, ,
\end{equation}
where $\vec\sigma$ is (double) the spin of the nucleon; $\vec{q}$ and $\vec{q}^{\, \prime}$ are the centre-of-mass (CM) $3$-momenta of the incoming and outgoing pions, respectively.

Equation (\ref{eq:EQ001}) defines the isoscalar (subscript $0$) and isovector (subscript $1$) $s$-wave scattering lengths ($b_0$ and $b_1$) and $p$-wave scattering volumes ($c_0$, $c_1$, $d_0$, and $d_1$), which may be 
projected onto the spin-isospin basis via standard transformations. The third term on the right-hand side of Eq.~(\ref{eq:EQ001}) is the no-spin-flip $p$-wave part of the $\pi N$ scattering amplitude, whereas the fourth 
term is the spin-flip part.

Also included in the low-energy $\pi N$ DB~\footnote{These measurements do not enter the SAID $\pi N$ DB.} are the two $\pi^- p$ scattering lengths $a_{cc} \equiv b_0 - b_1 $ and $a_{c0} \equiv \sqrt{2} b_1$ (corresponding 
to the $\pi^- p$ ES and CX reactions, respectively), obtained via the Deser~\footnote{I prefer this short form as reference to the two important relations developed by Deser, Goldberger, Baumann, and Thirring in 1954 
\cite{Deser1954}, as well as by Trueman in 1961 \cite{Trueman1961}, rather than Deser-Goldberger-Baumann-Thirring, Deser-Trueman, or Trueman-Deser relations.} formulae from measurements of the strong-interaction shift 
(henceforth, strong shift) $\epsilon_{1 s}$ \cite{Schroeder2001,Hennebach2014} and of the total decay width $\Gamma_{1s}$ \cite{Schroeder2001,Hirtl2021} of the ground state in pionic hydrogen~\footnote{The publication of 
the final result of the Pionic Hydrogen Collaboration ($\Gamma_{1s}=0.856(16)(22)$ eV \cite{Hirtl2021}) for the total decay width of the ground state in pionic hydrogen in 2021 reinforces my claim that all former values, 
disseminated by the collaboration (i.e., $\Gamma_{1s} \leq 850$ meV in 2003, $765(56)$ meV in 2009, ``from 750 to 900 meV with a central value at 850 meV'' in 2015, etc.), were preliminary and, as such, they should never 
have entered any analyses, inasmuch as the final result ($\Gamma_{1s}=0.868(40)(38)$ eV) of the former experiment \cite{Schroeder2001} was available since 1999. I see no reason why a preliminary result should be given 
priority over a final one, see also comments in Section 3.3 of Ref.~\cite{Matsinos2019}.}. The experiments were conducted at the Paul Scherrer Institut (PSI) during about one decade~\footnote{The results of a predecessor 
experiment, which had yielded compatible estimates (but larger uncertainties) for $\epsilon_{1 s}$ and $\Gamma_{1s}$, had already appeared in the mid 1990s \cite{Sigg1995,Sigg1996a}. The same collaboration also measured 
the strong shift and the total decay width of the ground state in pionic deuterium \cite{Chatellard1995,Chatellard1997}.} around the turn of the millennium. The quantities $a_{cc}$ and $a_{c0}$ represent the $\pi^- p$ ES 
and CX scattering amplitudes at $T = 0$ MeV, respectively.

A few words about the precision, attained by the PSI experiments, are in order. After the raw results (measured distributions of the $3 p \to 1 s$ transition energy) of the two experiments are corrected for the various 
Quantum-Electrodynamical effects reported in Ref.~\cite{Schlesser2011} (e.g., vacuum-polarisation effects, Breit-Pauli interaction, finite-size effects, self-energy effects, and so on), the two results for the strong shift 
$\epsilon_{1 s}$ \cite{Schroeder2001,Hennebach2014} come out nearly identical~\footnote{The corrected $\epsilon_{1s}$ result of Ref.~\cite{Schroeder2001} was given in Ref.~\cite{Schlesser2011} as: $-7.085(13)(34)$ eV, 
whereas the corresponding result of Ref.~\cite{Hennebach2014} (retaining the sign convention of Ref.~\cite{Schroeder2001}) was: $-7.0858(71)(64)$ eV. Both values ought to be corrected by $+0.0055$ eV, due to changes in 
some physical constants, namely in the charged-pion and proton rest masses, as well as in the charge radii of both particles.}. However, the perfect agreement between the two results is not the only reason to rejoice. 
Judged by Hadronic-Physics standards, the precision, attained in these experiments, is unequalled: the (relative) statistical uncertainty of $\epsilon_{1 s}$ in the former experiment is below the $2$ per-mille level, 
whereas the systematic uncertainty does not exceed $0.5~\%$; in case of the most recent experiment, both uncertainties are at the $1$ per-mille level, thus making the strong shift $\epsilon_{1 s}$ one of the best-known 
physical quantities in Hadronic Physics.

I find it unfortunate that the high precision at which $\epsilon_{1 s}$ is experimentally known does not translate into solid knowledge of the $\pi^- p$ ES length $a_{cc}$. Additional EM effects must be removed, and, as I 
will demonstrate in Section \ref{sec:ResidualEMAtThreshold}, the overall status of these corrections (at present) is below the expectations.

\section{\label{sec:Analysis}Extraction of the hadronic quantities from the data}

What does one need in order to attempt the description of the $\pi N$ data? The general approach comprises four steps. One needs to:
\begin{itemize}
\item[a)] model the hadronic part of the $\pi N$ interaction,
\item[b)] include the EM effects, 
\item[c)] select an input DB, and
\item[d)] select/establish an optimisation procedure.
\end{itemize}

A number of studies omit step (b) above, by fitting to the $s$- and $p$-wave $\pi N$ phase shifts of popular partial-wave analyses (PWAs), usually of those performed by the SAID group \cite{SAID}. One such solution 
in the recent past was their WI08 solution \cite{Arndt2006}, whereas their `current' solution has been named XP15 \cite{XP15}. The differences between these two solutions are small. Two comments are in order.
\begin{itemize}
\item Such a practice frequently gives rise to a misleading complacency: many authors and readers tend to forget that the problem of the EM corrections has been \emph{bypassed}, not resolved. Fitting to $\pi N$ phase shifts 
does not fix by itself any problems regarding the inclusion of the EM interaction; this issue cannot be relegated by simply changing the type of the input to a study.
\item Owing to the overconstrained fits, dispersion-relation analyses cannot provide meaningful estimates for the uncertainties of the $\pi N$ phase shifts; only their single-energy solutions are accompanied by such 
uncertainties. This implies that working uncertainties must be assigned in the studies which use the $\pi N$ phase shifts of the SAID solutions as input. On the contrary, all results obtained within the ETH $\pi N$ project 
(e.g., for the low-energy constants of the $\pi N$ interaction, for the $\pi N$ phase shifts, etc.) are accompanied by ($1 \sigma$) uncertainties which reflect the statistical and systematic fluctuations of the fitted data.
\end{itemize}

\subsection{\label{sec:Modelling}Modelling of the hadronic part of the $\pi N$ interaction}

The low-energy region corresponds to the non-perturbative domain of Quantum Chromodynamics (QCD), i.e., of the gauge field theory which describes the strong interaction between quarks and gluons. The largeness of the strong 
coupling constant $\alpha_S$ in this domain (which is also known as `confinement region') renders the framework of perturbation theory ineffective as far as the evaluation is concerned of the important physical characteristics 
of the various hadronic processes (i.e., of the scattering amplitudes and of the measurable quantities emanating thereof). To enable the modelling of the strong interaction between pions and nucleons, a number of approaches 
have been developed~\footnote{The reader must bear in mind that providing a complete list of references on the topic of this section lies well beyond the scope of this work. This subject is covered in dedicated review 
articles.}.
\begin{itemize}
\item Several approaches are based on dispersion relations. Information on the various forms of such schemes (fixed-$t$, forward, hyperbolic dispersion relations, etc.) may be found in Section 6 of H{\"o}hler's celebrated 
work \cite{Hoehler1983}. This category also includes the recent attempts to account for the data on the basis of the Roy-Steiner equations \cite{Ditsche2012,Hoferichter2016}.
\item Several effective-field models have emerged, inspired by the Chiral-Perturbation Theory ($\chi$PT) \cite{Leutwyler1994} in its various forms, e.g., Heavy Baryon $\chi$PT (HB$\chi$PT) \cite{Jenkins1991,Bernard1992}, 
Covariant Baryon $\chi$PT (CB$\chi$PT) \cite{Alarcon2013,Yao2016,Siemens2017}, etc. A Lorentz-invariant formulation of the Baryon $\chi$PT, featuring the method of infrared regularisation as the means to preserve the 
power-counting features of $\chi$PT at low energy, surfaced in Ref.~\cite{Becher1999}. Additional works are listed in Ref.~\cite{Huang2020} (as well as in a variety of other studies).
\item Several models, based on hadronic exchanges, were put forward during the last half century or so. Most of these efforts were critically examined in Section 6.4 of Ref.~\cite{Goudsmit1994}; one additional model 
\cite{Pascalutsa2000} appeared after Ref.~\cite{Goudsmit1994} was published. As aforementioned, the history of the development of the hadronic model, which is the backbone of the ETH $\pi N$ project, may be found online 
\cite{Matsinos2017a}.
\item Several phenomenological/empirical models have appeared, which are based on hadronic potentials \cite{Gibbs1995}, on the polynomial parameterisation of the $s$- and $p$-wave $K$-matrix elements \cite{Fettes1997,Matsinos2022a}, 
on the parameterisation of the $s$- and $p$-wave phase shifts \cite{Gibbs1998,Gibbs2005}, etc.
\end{itemize}

All aforementioned methods model the hadronic part of the $\pi N$ interaction on the basis of some parameters. (Some $\chi$PT-based approaches also model the EM part of the $\pi N$ interaction.) The methods of the first two 
categories are more constrained, in that they fulfil the theoretical constraints of unitarity (see Section 5 of Ref.~\cite{Hoehler1983}), analyticity (see Section 7 of Ref.~\cite{Hoehler1983}), crossing symmetry~\footnote{The 
scattering amplitudes of the two ES reactions are linked via the interchange $s \leftrightarrow u$ in the two invariant amplitudes $A_{\pm}(s,t,u)$ and $B_{\pm}(s,t,u)$, where $s$, $t$, and $u$ are the standard Mandelstam 
variables.} (see Section 6.1 of Ref.~\cite{Hoehler1983}), and isospin invariance (see Section 2.2 of Ref.~\cite{Hoehler1983}, as well as the historical perspective of Ref.~\cite{Rasche1971} on the concept of isospin). The 
partial-wave amplitudes of the hadronic model of Ref.~\cite{Matsinos2017a} fulfil unitarity, crossing symmetry, and isospin invariance. On the other hand, the phenomenological/empirical models are compatible only with the 
unitarity constraint. This remark, however, must not be taken as indicating a disadvantage; being devoid of the theoretical constraints of crossing symmetry and isospin invariance, these data-driven approaches
\begin{itemize}
\item grant maximal freedom to the fitted data and
\item are indispensable tools in the tests of these theoretical constraints at low energy.
\end{itemize}

Given its relevance in this work, a few words about the isospin invariance are in order. Assuming that the isospin invariance holds in the hadronic part of the $\pi N$ interaction, only two (complex) scattering amplitudes 
enter the description of the three low-energy $\pi N$ reactions (e.g., see Appendix 1 in Ref.~\cite{Matsinos1997}): the isospin $I=3/2$ amplitude ($f_3$) and the $I=1/2$ amplitude ($f_1$). Fulfilment of the isospin 
invariance in the hadronic part of the $\pi N$ interaction implies that the $\pi^+ p$ reaction is described by $f_3$, the $\pi^- p$ ES reaction by the linear combination $(f_3 + 2 f_1)/3$, and the $\pi^- p$ CX reaction by 
$\sqrt{2} (f_3-f_1)/3$. From these relations, the following expression (known as `triangle identity') links together the amplitudes $f_{\pi^+ p}$, $f_{\pi^- p}$, and $f_{\rm CX}$:
\begin{equation} \label{eq:EQ002}
f_{\pi^+ p} - f_{\pi^- p} = \sqrt{2} f_{\rm CX} \, \, \, .
\end{equation}
In the following, `isospin invariance in the $\pi N$ interaction' will be used as the short form of `isospin invariance in the hadronic part of the $\pi N$ interaction'; it is known that the isospin invariance is broken in 
the EM part of the interaction. One the proposed indicators of the violation of the isospin invariance is the symmetrised relative difference $R_2$:
\begin{equation} \label{eq:EQ002_5}
R_2 \coloneqq 2 \frac{\Re \left[ f_{\rm CX} - f^{\rm extr}_{\rm CX} \right]}{\Re \left[ f_{\rm CX} + f^{\rm extr}_{\rm CX} \right]} =
 2 \frac{ \Re \left[ f_{\pi^+ p} - f_{\pi^- p} - \sqrt{2} f^{\rm extr}_{\rm CX} \right] }{ \Re \left[ f_{\pi^+ p} - f_{\pi^- p} + \sqrt{2} f^{\rm extr}_{\rm CX} \right] } \, \, \, ,
\end{equation}
where the operator $\Re$ returns the real part of a complex number and the $f^{\rm extr}_{\rm CX}$ is extracted from the data (as opposed to the reconstructed amplitude $f_{\rm CX}$, obtained via Eq.~(\ref{eq:EQ002})).

\begin{figure}
\begin{center}
\includegraphics [width=15.5cm] {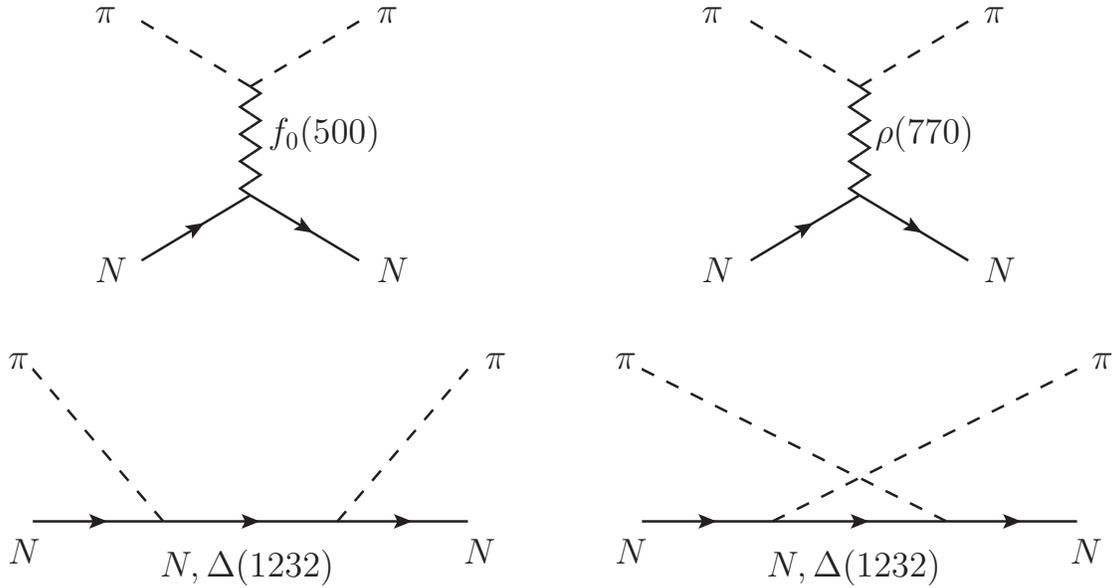}
\caption{\label{fig:FeynmanGraphsETHZ}The main Feynman diagrams of the ETH model: scalar-isoscalar ($I=J=0$) and vector-isovector ($I=J=1$) $t$-channel graphs (upper part), and $N$ and $\Delta(1232)$ $s$- and $u$-channel 
graphs (lower part). Not shown in this figure, but also analytically included in the model, are the small contributions from the well-established (four-star) $s$ and $p$ higher baryon resonances with masses below $2$ GeV 
and known branching fractions to $\pi N$ decay modes, as well as those from the $t$-channel exchanges of four (three scalar-isoscalar and one vector-isovector) mesons with masses below $2$ GeV and known branching fractions 
to $\pi \pi$ decay modes, see Ref.~\cite{Matsinos2017a} for details.}
\vspace{0.25cm}
\end{center}
\end{figure}

Of interest in this work are those of the approaches which can be used as tools to test the theoretical constraint of the isospin invariance in the $\pi N$ interaction. This personal predilection is the product of the 
experience I gained from the analysis of the low-energy $\pi N$ DB during the last three decades. In retrospect, the separate analysis of the measurements of the three low-energy $\pi N$ reactions (or the joint analyses of 
the measurements after leaving out one of these reactions), has given rise to persistent discrepancies. In this respect, of interest in this work are the models of Refs.~\cite{Matsinos2017a,Gibbs1995,Fettes1997,Matsinos2022a,Matsinos1997}. 
It is interesting to examine the similarities of and the differences between the aforementioned models in the chronological order in which the first analyses, involving these models, appeared.
\begin{itemize}
\item In their pioneering work \cite{Gibbs1995}, Gibbs and collaborators extracted the $\pi N$ scattering amplitudes $f_{\pi^+ p}$, $f_{\pi^- p}$, and $f^{\rm extr}_{\rm CX}$ from the (sparse, at that time) low-energy $\pi N$ 
DB using hadronic potentials of different shapes. Subsequently, the authors evaluated the difference between the extracted $f^{\rm extr}_{\rm CX}$ and the reconstructed $f_{\rm CX}$ scattering amplitudes, and established a 
significant discrepancy in the $s$-wave part of the interaction (see their Fig.~1); a smaller discrepancy was reported in the no-spin-flip $p$-wave part of the interaction (see their Fig.~2). These discrepancies were found 
to be independent (on the whole) of the type of the hadronic potential used in the data description (see their Table 1). Another interesting result in that paper is frequently overlooked: the zero-crossings of the extracted 
$f^{\rm extr}_{\rm CX}$ and reconstructed $f_{\rm CX}$ scattering amplitudes (a phenomenon which is due to the $s$- and $p$-wave destructive interference at forward CM scattering angles $\theta$) occur at different energies 
(see their Fig.~3). Subsequent studies \cite{Matsinos2017a,Matsinos1997}, using sizeably larger (in comparison with Ref.~\cite{Gibbs1995}) DBs, corroborated these results~\footnote{The only outcome of the analyses, performed 
by Gibbs and collaborators, which cannot be corroborated by the results obtained within the ETH $\pi N$ project, is the equality (within the uncertainties) of the two $p$-wave phase shifts $\delta^{3/2}_{1-}$ ($P_{31}$) and 
$\delta^{1/2}_{1+}$ ($P_{13}$) \cite{Gibbs1998}.}. This is more remarkable than it sounds because, for the sake of argument, the $\pi^- p$ CX DB has been increased from $46$ to $337$ datapoints since the time the analysis 
of Ref.~\cite{Gibbs1995} was performed.
\item When Fettes and I sought in the mid 1990s the description of the low-energy $\pi^+ p$ data on the basis of simple polynomial forms for the $s$- and $p$-wave $K$-matrix elements \cite{Fettes1997}, I could not foresee 
the pivotal role, which this novel model-independent way of analysis would assume within the ETH $\pi N$ project in the years to follow. This is due to two reasons: first, the method has been firmly established as the means 
of identification of the outliers which the low-energy $\pi N$ DB contains; second, it provides a model-independent way of analysing the measurements, one which is devoid of theoretical constraints. There are two issues in 
analyses using similar parameterisation:
\begin{enumerate}
\item the number of terms which one retains from the power series (in a suitable variable, e.g., in the pion CM kinetic energy $\epsilon$ or, more frequently, in the square of the CM $3$-momentum in the initial state 
\cite{Matsinos2022a}), and
\item the forms which one uses in the modelling of the resonant contributions.
\end{enumerate}
Experience has shown that the polynomial parameterisation of the $s$- and $p$-wave $K$-matrix elements successfully captures the dynamics of the $\pi N$ interaction at low energy. A number of tests have been performed, 
demonstrating beyond doubt that the outliers are flanked by measurements which can successfully be accounted for. Therefore, the identification of datapoints as outliers in the fits cannot be attributed to the inadequacy 
of the employed parametric forms to account for the energy dependence of the $\pi N$ phase shifts; consequently, the outliers are suggestive of experimental shortcomings. In the data analysis, terms up to 
$\mathcal{O}(\epsilon^2)$ (or up to $\mathcal{O}(q^4)$ in the parameterisation of the $K$-matrix elements in powers of $q^2$, where $q \coloneqq \lvert \vec{q} \, \rvert$) are retained \cite{Matsinos2022a}. Owing to the 
current uncertainties of the measurements, the coefficients of higher orders cannot be determined reliably from the data.
\item The backbone of the ETH $\pi N$ project is the isospin-invariant hadronic model whose Feynman diagrams are shown in Fig.~\ref{fig:FeynmanGraphsETHZ}. The question, of course, arises as to how an isospin-invariant model 
can be relevant in the tests of the isospin invariance; in fact, there are two ways by which it can. One approach rests upon the extraction of a phase-shift solution from the ES DBs and the subsequent evaluation of the 
scattering amplitude, corresponding to the $\pi^- p$ CX reaction, using Eq.~(\ref{eq:EQ002}) \cite{Matsinos1997}. Having reconstructed the amplitude $f_{\rm CX}$, one may generate predictions for the various observables, 
corresponding to the kinematical quantities ($T$,$\theta$) of the measurements in the $\pi^- p$ CX DB. The comparison between these predictions and the experimental data enables the assessment of the amount by which the 
triangle identity (equivalently, the isospin invariance) is violated; this method does not require the extraction of the amplitude $f^{\rm extr}_{\rm CX}$ from the data. A second approach was more recently established 
\cite{Matsinos2017a}. Two kinds of fits may be pursued: one uses as input DB the data of two ES reactions, whereas the second uses the data of the $\pi^+ p$ and $\pi^- p$ CX reactions. In both cases, the isospin $I=3/2$ 
partial-wave amplitudes are fixed (predominantly) from the $\pi^+ p$ ES reaction, whereas the two $\pi^- p$ reactions establish the $I=1/2$ partial-wave amplitudes. The differences between the two phase-shift solutions (as 
well as those between the two sets of fitted values of the model parameters~\footnote{When an isospin-invariant model is used in the description of data which contain isospin-breaking effects (such as the measurements of 
the $\pi^- p$ CX reaction presumably do), the isospin-breaking effects are absorbed in the fitted values of the model parameters, which thus become effective. Such changes have been observed in the joint fits of the ETH 
model to the $\pi^+ p$ and $\pi^- p$ CX data: affected are the Fermi-like coupling associated with the $t$-channel $\rho$-meson-exchange Feynman diagram (model parameter $G_\rho$) and (to a lesser degree) the $\pi N$ 
coupling constant (model parameter $g_{\pi N N}$) \cite{Matsinos2017a}. The result of the presence of effects in the data, which have no counterpart in the modelling of the hadronic part of the $\pi N$ interaction, is the 
inevitable increase in the $\chi^2$ value corresponding to the optimal description of the input DB.}) may yield a measure of the departure from the triangle identity.
\end{itemize}

It is true that those of the approaches, which fix their parameters from fits to popular phase-shift solutions, could (in principle) test the isospin invariance in the $\pi N$ interaction, e.g., by using as input the 
$\pi N$ phase shifts obtained from specific reactions or combinations of reactions (e.g., two such sets of $\pi N$ phase shifts are generated within the ETH $\pi N$ project). However, this has never been done; all such 
works choose to make use of $\pi N$ phase shifts which have been obtained by global fits to the $\pi N$ data (see end of Section \ref{sec:Optimisation}).

\subsection{\label{sec:EMCorrections}Inclusion of the EM effects} 

The EM effects come in two forms:
\begin{itemize}
\item as direct contributions to the $\pi N$ scattering amplitude (Coulomb amplitude) for the two ES reactions and
\item as distortion effects on the $\pi N$ phase shifts and partial-wave amplitudes, originating from the interference of the EM and strong interactions.
\end{itemize}

The first systematic attempt to determine the EM corrections, applicable to the data in the scattering region ($T>0$ MeV), was performed in the 1970s, and led to three papers by the NORDITA group 
\cite{Tromborg1976,Tromborg1977,Tromborg1978}. Their results are listed in tabular form (for several CM momenta) spanning a broad energy domain. The NORDITA programme did not develop a procedure for correcting any 
measurements at the $\pi N$ threshold, though (not known to some) the issue of the EM corrections at the $\pi N$ threshold had already been addressed \cite{Oades1971a,Oades1971b,Rasche1976a,Rasche1976b} a few years prior 
to the publication of the NORDITA papers. In Ref.~\cite{Rasche1976b}, Rasche and Woolcock developed the methodology for the inclusion of the effects of the $\pi^- p \to \gamma n$ channel (simply $\gamma n$ channel 
henceforth) at the $\pi N$ threshold, whereas the same authors presented in a subsequent paper \cite{Rasche1982} a method for the extraction of the (corrected) strong shift and total decay width in pionic atoms. The 
numerical results in Section 3 of Ref.~\cite{Rasche1982} were tailored to the $2 p \to 1 s$ transition in pionic hydrogen, which was considered (on the basis of the yield) the most promising transition at that time. Comments 
on the corrections of Ref.~\cite{Rasche1982} may be found in Section 4 of Ref.~\cite{Sigg1996b}. The corrections of Ref.~\cite{Rasche1982} were superseded by those of Ref.~\cite{Oades2007}.

The inclusion of the EM effects in the scattering region for $T < 300$ MeV was also addressed by Bugg in 1973 \cite{Bugg1973}: these `Coulomb-barrier' corrections, determined by solving a relativised Schr{\"o}dinger 
equation, were used in many analyses of the $\pi N$ data at former times.

To my knowledge, the only programme which addressed the issue of the EM corrections throughout the low-energy region (also including, in a consistent manner, the corrections which ought to apply to the measurements of the 
$\epsilon_{1 s}$ and $\Gamma_{1s}$ in pionic hydrogen) was carried out at the University of Zurich in the late 1990s and the early 2000s. The programme resulted in the publication of the EM corrections for $\pi^+ p$ 
scattering \cite{Gashi2001a}, for $\pi^- p$ scattering \cite{Gashi2001b}, as well as for the strong shift $\epsilon_{1 s}$ and the total decay width $\Gamma_{1s}$ \cite{Oades2007}. The programme rested upon the determination 
of the hadronic potentials which optimally account for the low-energy $\pi N$ DB in the scattering region, and made use of the same potentials to determine the corrections at the $\pi N$ threshold using a three-channel 
($\pi^- p \to \pi^- p, \pi^0 n, \gamma n$) calculation. Unfortunately, this method is applicable only at low energy.

Some of the available schemes \cite{Gibbs1998,Tromborg1976,Tromborg1977,Tromborg1978,Gashi2001a,Gashi2001b} for the determination of the EM corrections for $\pi N$ scattering experiments are compared in Ref.~\cite{Gibbs2005}. 
Two conclusions may be drawn from that work: first, that the effects of the EM corrections on the low-energy $\pi N$ DCSs cannot be neglected; second, that the differences among the available schemes are generally small. The 
second realisation is encouraging for analysts of the low-energy $\pi N$ data, in that the selection of the EM-correction scheme seems to be of lesser importance than applying no EM corrections to the $\pi N$ phase shifts 
and partial-wave amplitudes, obtained from the modelling of the hadronic part of the $\pi N$ interaction.

To summarise, no consistent approach exists for the treatment of the EM corrections in the entirety of the energy range of the available data, i.e., from the $\pi N$ threshold to the few-GeV region.

\subsection{\label{sec:Optimisation}Selection of an analysis method/optimisation scheme}

In Particle Physics, the commonest (by far) choice in optimisation procedures is the minimisation of $\chi^2$-based functions. Of interest herein is the $\chi^2$-based minimisation with rescaling of the input datasets 
\cite{Arndt1972}, though also $\chi^2$-based methods without rescaling \cite{Fettes1997}, as well as more robust methods \cite{Matsinos1997}, have been used in the analysis of the low-energy $\pi N$ data. Some in-between 
methods, listed in Section 4.1 of Ref.~\cite{Matsinos2019}, could be tested in the future. In any case, by variation of the parameters of the various hadronic models, one achieves the optimisation of the description of the 
input data.

The PSAs of the $\pi N$ data by the SAID group, as well as those performed within the ETH $\pi N$ project during the last two decades, make use of the minimisation function which Arndt and Roper introduced half a century ago 
\cite{Arndt1972}. The contribution of the $j$-th dataset to the overall $\chi^2$ reads as:
\begin{equation} \label{eq:EQ003}
\chi^2_j=\sum_{i=1}^{N_j} \left( \frac{y_{ij}^{\rm exp} - z_j y_{ij}^{\rm th}}{\delta y_{ij}^{\rm exp} } \right)^2 + \left( \frac{z_j-1}{\delta z_j} \right)^2 \, \, \, ,
\end{equation}
where $y_{ij}^{\rm exp}$ denotes the $i$-th datapoint of the $j$-th dataset, $y_{ij}^{\rm th}$ the corresponding fitted (`theoretical') value, $\delta y_{ij}^{\rm exp}$ the statistical uncertainty of $y_{ij}^{\rm exp}$, 
$z_j$ a scale factor (which applies to the entire dataset), $\delta z_j$ the normalisation uncertainty (reported or assigned), and $N_j$ the number of the accepted datapoints in the dataset (i.e., of the datapoints which 
remain in the dataset after the removal of any outliers). At each optimisation step, the fitted values $y_{ij}^{\rm th}$ are supplied by the modelling of the hadronic part of the $\pi N$ interaction and the inclusion of 
the EM effects. As the scale factor $z_j$ appears only in $\chi^2_j$, the minimisation of the overall $\chi^2 \coloneqq \sum_{j=1}^{N} \chi^2_j$ (where $N$ denotes the number of the datasets in the fit) implies the fixation 
of each scale factor $z_j$ from the condition $\partial \chi^2_j / \partial z_j = 0$. The unique solution
\begin{equation} \label{eq:EQ004}
z_j = \frac{\sum_{i=1}^{N_j} y_{ij}^{\rm exp} y_{ij}^{\rm th} / (\delta y_{ij}^{\rm exp} )^2 + (\delta z_j)^{-2}} {\sum_{i=1}^{N_j} (y_{ij}^{\rm th} / \delta y_{ij}^{\rm exp})^2 + (\delta z_j)^{-2}}
\end{equation}
leads to
\begin{equation*}
(\chi^2_j)_{\rm min} = \sum_{i=1}^{N_j} \frac{ (y_{ij}^{\rm exp} - y_{ij}^{\rm th})^2}{(\delta y_{ij}^{\rm exp})^2 } - \frac {\left( \sum_{i=1}^{N_j} (y_{ij}^{\rm exp} - y_{ij}^{\rm th}) y_{ij}^{\rm th} / (\delta y_{ij}^{\rm exp} )^2 \right)^2} 
{ \sum_{i=1}^{N_j}(y_{ij}^{\rm th}/\delta y_{ij}^{\rm exp})^2 + (\delta z_j)^{-2} } \, \, \, .
\end{equation*}
The sum of the contributions $\sum_{j=1}^{N} (\chi^2_j)_{\rm min}$ is a function of the parameters entering the modelling of the hadronic part of the $\pi N$ interaction. By variation of these parameters, the overall $\chi^2$ 
is minimised, yielding $\chi^2_{\rm min}$.

The numerical minimisation may be achieved by using many commercial and non-commercial software libraries. Within the ETH $\pi N$ project, the MINUIT package \cite{James} of the CERN library (FORTRAN version) has exclusively 
been used. Each optimisation is achieved by following the sequence: SIMPLEX, MINIMIZE, MIGRAD, and MINOS. The calls to the last two methods involve the high-level strategy in the numerical minimisation.
\begin{itemize}
\item SIMPLEX uses the simplex method of Nelder and Mead \cite{Nelder1965}.
\item MINIMIZE calls MIGRAD, but reverts to SIMPLEX in case that MIGRAD fails to converge.
\item MIGRAD, undoubtedly the workhorse of the MINUIT software package, is a variable-metric method, based on the Davidon-Fletcher-Powell algorithm. The method checks for the positive-definiteness of the Hessian matrix.
\item MINOS performs a detailed error analysis, separately for each model parameter, taking into account the non-linearities in the problem, as well as the correlations among the parameters.
\end{itemize}
All aforementioned methods admit an optional argument, fixing the maximal number of calls to each method (separately). If this limit is reached, the corresponding method is terminated (by MINUIT, internally) regardless of 
whether or not that method converged. Inspection of the (copious) MINUIT output may easily ascertain whether or not the application terminated successfully and its methods converged.

Before entering the details of the analysis of the $\pi N$ data, I will outline the types of fits which one may pursue.
\begin{itemize}
\item In the \emph{global} fits, all measurements of the three experimentally-accessible $\pi N$ reactions at low energy comprise the input DB, whose description ought to be optimised. Global fits to the $\pi N$ data are 
routinely performed by the SAID group. Not known to some, global fits to the low-energy $\pi N$ data are possible within the ETH $\pi N$ project since the mid 1990s. Such fits are carried out (for the sake of completeness), 
though (due to reasons which will become clear later on) the results have never been reported/used.
\item In the \emph{joint} fits, measurements of two $\pi N$ reactions comprise the input DB. Such fits are performed within the ETH $\pi N$ project, using either the ETH model or the polynomial parameterisation of the $s$- 
and $p$-wave $K$-matrix elements. At present, two such fits are carried out: the first uses as input DB the ES data, whereas the second imports the data of the $\pi^+ p$ and $\pi^- p$ CX reactions. The third combination of 
reactions, i.e., the selection as input of the measurements contained in the DBs of the two $\pi^- p$ reactions (ES and CX), remains a possibility; such fits have not been carried out yet.
\item In the \emph{separate} fits, measurements of only one $\pi N$ reaction comprise the input DB. Such fits may be performed using the phenomenological/empirical models of Section \ref{sec:Modelling} 
\cite{Gibbs1995,Fettes1997,Matsinos2022a,Gibbs1998,Gibbs2005}. Due to the sizeable correlations among the model parameters, separate fits of the ETH model to the $\pi N$ data have not been attempted for a long time.
\end{itemize}

\section{\label{sec:Lessons}So, what has been learnt?}

The studies, which do not pursue global fits to the low-energy $\pi N$ data (and can therefore test the isospin invariance in the $\pi N$ interaction), are those performed by Gibbs and collaborators \cite{Gibbs1995}, 
as well as the ones carried out within the ETH $\pi N$ project since the mid 1990s (see cited works in Ref.~\cite{Matsinos2017a}). In spite of their substantial differences in the modelling of the hadronic part of the $\pi N$ 
interaction, in the inclusion of the EM effects, in the input DB (e.g., the low-energy $\pi N$ DB grew substantially after Ref.~\cite{Gibbs1995} appeared), and in the choice of the optimisation procedure, the results of 
these works strongly suggest that, regarding the $\pi N$ interaction at low energy, not all pieces fall into place. Reference \cite{Gibbs1995} demonstrated that the discrepancy in the $s$ wave between the extracted $f^{\rm extr}_{\rm CX}$ 
and the reconstructed $f_{\rm CX}$ scattering amplitudes is statistically significant between $T=30$ and $50$ MeV, e.g., see Fig.~1 therein. Given the growing importance of the $p$-wave part of the $\pi N$ scattering 
amplitude with increasing energy, Ref.~\cite{Gibbs1995} essentially predicts (though not explicitly mentioned in the paper) an energy-dependent isospin-breaking effect in the DCSs of the $\pi^- p$ CX reaction.

Within the ETH $\pi N$ project, such an effect has been established \cite{Matsinos2017a}. Integrated between $0$ and $100$ MeV, the discrepancy between measured (or extracted) and predicted (or reconstructed) cross sections 
would be equivalent to an effect at the level of $(12.8 \pm 1.3) \cdot 10^{-2}$ or, naively converted into a relative difference between the two $\pi^- p$ CX scattering amplitudes, of $r_2=6.19(63)~\%$. The aforementioned 
discrepancy between measured and predicted cross sections indicates that
\begin{equation*}
\avg{\frac{\lvert f^{\rm extr}_{\rm CX} \rvert^2}{\lvert f_{\rm CX} \rvert^2}} > 1 \, \, \, ,
\end{equation*}
where
$f_{\rm CX}$ is obtained from $f_{\pi^+ p}$ and $f_{\pi^- p}$ via Eq.~(\ref{eq:EQ002}). As a result,
\begin{equation*}
\avg{\frac{\lvert f^{\rm extr}_{\rm CX} \rvert^2}{\lvert f_{\rm CX} \rvert^2}} = (1 + r_2)^2 \, \, \, ,
\end{equation*}
for some $r_2 \in \mathbb{R}_{>0}$. To obtain a working (though, given the pronounced energy dependence of the effect, not mathematically meaningful) $R_2$ result, one could assume the constancy of the ratio on the left-hand 
side of the previous equation, in which case
\begin{equation} \label{eq:EQ005}
\frac{\lvert f^{\rm extr}_{\rm CX} \rvert}{\lvert f_{\rm CX} \rvert} = \sqrt{2} \frac{\lvert f^{\rm extr}_{\rm CX} \rvert}{\lvert f_{\pi^+ p} - f_{\pi^- p} \rvert} = (1 + r_2) \, \, \, .
\end{equation}
At low energy, the imaginary parts of the scattering amplitudes may be omitted, and one obtains
\begin{equation*}
\frac{\Re \left[ f^{\rm extr}_{\rm CX} \right] }{\Re \left[ f_{\pi^+ p} - f_{\pi^- p} \right] } \approx \frac{1 + r_2}{\sqrt{2}} \, \, \, .
\end{equation*}
Using Eq.~(\ref{eq:EQ002_5}), one ends up with the following relation for $r_2 \ll 1$.
\begin{equation*}
R_2 \approx \frac{-2 r_2}{2 + r_2} = -r_2 + \frac{r_2^2}{2} - \frac{r_2^3}{4} + \dots
\end{equation*}
Therefore, $R_2 \approx - r_2$ to a first approximation (the real part of the scattering amplitude $f_{\rm CX}$, which is obtained from the two ES scattering amplitudes via the triangle identity, is \emph{less} negative than 
$f^{\rm extr}_{\rm CX}$ at low energy); retaining all higher-order contributions, $R_2 \approx -6.01(59)~\%$. Although these percentages should not to be taken too seriously, they agree on the whole with the findings of 
Ref.~\cite{Gibbs1995}. Of course, the indicator $R_2$ of Eq.~(\ref{eq:EQ002_5}) has been obtained in other analyses at several energies, separately for the $s$ and $p$ waves, i.e., not in the form of an overall relative 
difference (which is how the discrepancy is currently presented within the ETH $\pi N$ project). In future reports relating to the ETH $\pi N$ project, the indicator $R_2$ will be evaluated at a few energies, separately 
for the $s$ and $p$ waves.

A few words on the stability of the results of Refs.~\cite{Matsinos2017a,Gibbs1995,Matsinos1997} are worth including in this preprint. To my knowledge, the results of Ref.~\cite{Gibbs1995} have never been updated. A relative 
deviation from the triangle identity (in the $s$ and $p$ waves, average value over three energies) had been obtained in Ref.~\cite{Matsinos1997}: $(6.4 \pm 1.4)~\%$. Between 1997 and 2019, only the $\chi^2$ value (and the 
resulting p-value) of the reproduction of the low-energy $\pi^- p$ CX data on the basis of the solution, obtained from the fits of the ETH model to the ES data, was used as a measure of the departure from the triangle 
identity. The quantity $r_2$ of Eq.~(\ref{eq:EQ005}) was first evaluated in 2019 ($6.17(64)~\%$), and then updated in 2020 ($6.19(63)~\%$). Owing to the enhancement of the low-energy $\pi^- p$ CX DB over the past twenty-five 
years, the statistical significance of the departure from the triangle identity improved over time, though the amount of the departure itself has remained nearly unchanged.

The outcome of the optimisation within the ETH $\pi N$ project is routinely subjected to further analysis. As aforementioned, two kinds of fits are pursued: one uses as input DB the data of two ES reactions, the other the 
data of the $\pi^+ p$ and $\pi^- p$ CX reactions. Given that the Arndt-Roper formula \cite{Arndt1972}, see Eq.~(\ref{eq:EQ003}), is used in the optimisation, the expectation is that the datasets which are scaled `upwards' 
($z<1$) balance (on average) those which are scaled `downwards' ($z>1$) in both cases. Furthermore, the energy dependence of the scale factors $z_j$ of Eq.~(\ref{eq:EQ004}) must not be significant. If these prerequisites 
are not fulfilled, the description of the input DB cannot be considered satisfactory. To cut a long story short, the fitted values of the scale factor $z$ in the former case (fit to the ES data) do not show any significant 
departure from the statistical expectation (i.e., they are centred on $1$ and exhibit no significant energy dependence). On the other hand, significant systematic effects have been established in the latter case (fit to the 
data of the $\pi^+ p$ and $\pi^- p$ CX reactions) throughout the time this analysis has been performed: the modelling of the hadronic part of the $\pi N$ interaction at low energy generates (on average) \emph{overestimated} 
DCS values for the $\pi^+ p$ reaction and \emph{underestimated} ones for the $\pi^- p$ CX reaction.

To summarise, Refs.~\cite{Matsinos2017a,Gibbs1995,Matsinos1997} essentially demonstrate that the $\pi N$ scattering amplitudes $f_{\pi^+ p}$, $f_{\pi^- p}$, and $f^{\rm extr}_{\rm CX}$, extracted from the low-energy $\pi N$ 
DB, fail to satisfy the triangle identity of Eq.~(\ref{eq:EQ002}). Henceforth, I will refer to this failure as the `low-energy $\pi N$ enigma'. This discrepancy suggests that \emph{at least one} of the following assumptions 
is not valid.
\begin{itemize}
\item[a)] Any systematic effects (e.g., systematic underestimation/overestimation) in the determination of the absolute normalisation of the datasets of the low-energy $\pi N$ DB are not significant. The absolute 
normalisation of each dataset is subject only to statistical fluctuation, regulated by the (reported or assigned) normalisation uncertainty of that dataset.
\item[b)] Any residual contributions to the EM corrections of Refs.~\cite{Gibbs1998,Tromborg1976,Tromborg1977,Tromborg1978,Oades2007,Gashi2001a,Gashi2001b} are not significant.
\item[c)] The isospin invariance, i.e., Eq.~(\ref{eq:EQ002}), holds in the hadronic part of the $\pi N$ interaction.
\end{itemize}
The three possibilities, arising from the non-fulfilment of each of these assumptions, will be examined in Section \ref{sec:Explanations}. Before that, I ought to address two important questions.

\subsection{\label{sec:JointFits}Can global fits to all low-energy $\pi N$ data be performed?}

The answer to this question is: without doubt. A more meaningful question, however, would have been: are such fits satisfactory? Had I been asked that question, I would have replied without hesitation: hardly so. The 
low-energy $\pi N$ enigma cannot be swept under the carpet by submitting the available low-energy data to a global fit; one way or another, the discrepancy is bound to find its way to the output. In this section, I will 
examine one way by which this comes about, using the current solution of the SAID website. I ought to emphasise that the only input (the fitted values of the scale factor $z$ are available to two decimal places online 
\cite{SAID}) in this part of the study comes from the SAID website (copied on March 31, 2022); there is no input from any other source. Their $T_j$ and $\delta z_j$ values will be used.

Following an analysis of the energy dependence of the scale factors $z_j$, a systematic bias was established in Ref.~\cite{Matsinos2017b} in the output of the SAID solution WI08 \cite{Arndt2006}. Herein, I will re-examine 
the issue using the more recent SAID solution XP15 \cite{XP15}. There is no difference between the two sets of fitted values of the scale factor $z$ for the ES datasets, whereas small changes can be seen in some $\pi^- p$ 
CX datasets in the region of the $s$- and $p$-wave destructive interference, namely in the scale factors $z_j$ of (some of) the FITZGERALD86 datasets~\footnote{The FITZGERALD86 measurements of the $\pi^- p$ CX DCS in the 
kinematical region of the $s$- and $p$-wave interference minimum are arranged into seven three-point datasets. A fourth value was obtained in Ref.~\cite{Fitzgerald1986} via the extrapolation to $\theta = 0^\circ$ of a 
quadratic form fitted to the three datapoints of each dataset. The SAID $\pi N$ DB includes these additional datapoints, though (technically) they do not constitute independent observations. I do not insinuate that the 
addition of these seven datapoints is of any significance \emph{in practice}.} \cite{Fitzgerald1986}.

\begin{figure}
\begin{center}
\includegraphics [width=15.5cm] {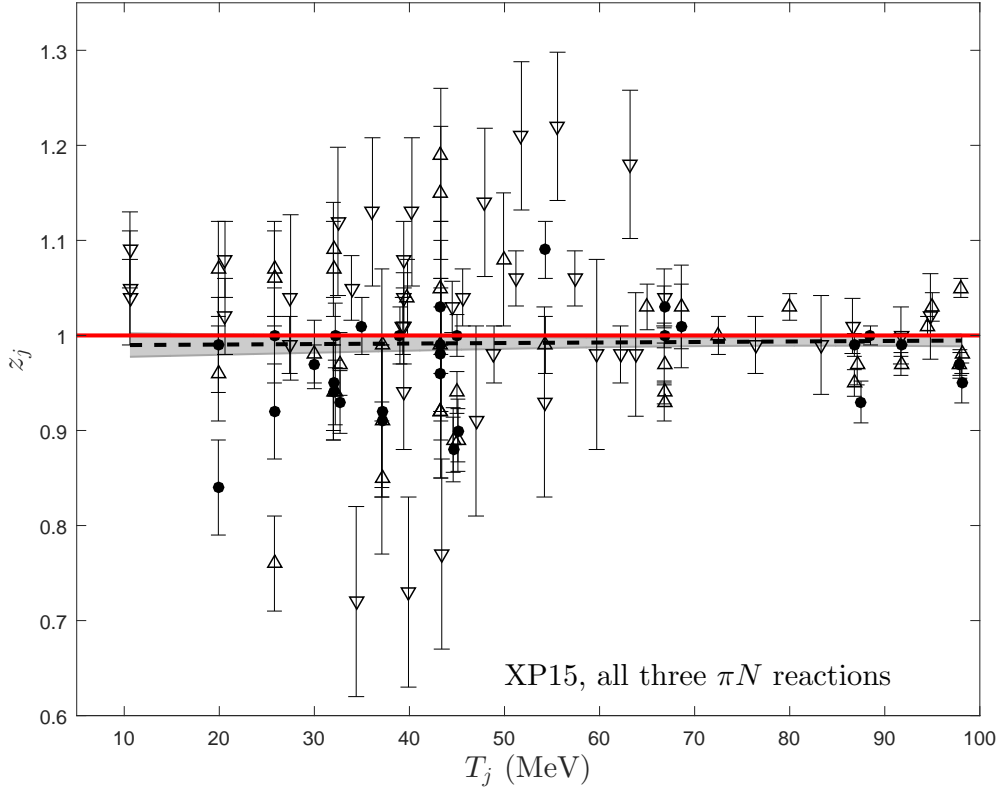}
\caption{\label{fig:ALLXP15}Plot of the scale factors $z_j$ of Eq.~(\ref{eq:EQ004}) corresponding to the XP15 solution \cite{XP15}. $T_j$ denotes the pion laboratory kinetic energy of the $j$-th dataset. The data shown 
correspond to DCSs only (upward triangles: $\pi^+ p$, filled circles: $\pi^- p$ ES, downward triangles: $\pi^- p$ CX). The datasets which were practically floated (the SAID group assign a normalisation uncertainty of 
$100~\%$ to datasets of suspicious absolute normalisation) have not been used. The dashed straight line represents the optimal, weighted linear fit to the data shown (see Table \ref{tab:XP15Parameters}), whereas the shaded 
band represents $1 \sigma$ uncertainties. The red line is the ideal outcome of the optimisation.}
\vspace{0.25cm}
\end{center}
\end{figure}

I will now report on a simple analysis of the scale factors $z_j$ in the XP15 solution for $T \leq 100$ MeV, starting from the results representing all available data of the three low-energy $\pi N$ reactions (see 
Fig.~\ref{fig:ALLXP15}). There is no doubt that, in spite of the large fluctuation which is present in the plot, the fitted values of the scale factor $z$ show no significant bias: their linear fit results in an intercept 
which is compatible with $1$ and a slope which is compatible with $0$, see Table \ref{tab:XP15Parameters}. Therefore, the conclusion at first glance is that their fit yields results which are compatible with the statistical 
expectation (i.e., with the unbiased outcome of an optimisation based on the Arndt-Roper formula \cite{Arndt1972}).

\vspace{0.5cm}
\begin{table}[h!]
{\bf \caption{\label{tab:XP15Parameters}}}The fitted values of the parameters of the weighted linear fit to the data shown in Figs.~\ref{fig:ALLXP15}-\ref{fig:PIMCXXP15}, as well as the fitted uncertainties, corrected with 
the application of the Birge factor $\sqrt{\chi^2/{\rm NDF}}$, which takes account of the goodness of each fit \cite{Birge1932}. Also quoted is the reduced $\chi^2$ value of each linear fit; NDF stands for the number of 
degrees of freedom.
\vspace{0.25cm}
\begin{center}
\begin{tabular}{|l|c|c|c|}
\hline
Reaction & Intercept & Slope ($10^{-3}$ MeV$^{-1}$) & $\chi^2$/NDF \\
\hline
\hline
All three $\pi N$ reactions & $0.989 \pm 0.014$ & $0.05 \pm 0.19$ & $355.45/108 \approx 3.29$ \\
\hline
$\pi^+ p$ & $0.936 \pm 0.028$ & $0.70 \pm 0.34$ & $167.53/39 \approx 4.30$ \\
$\pi^- p$ ES & $0.972 \pm 0.021$ & $0.16 \pm 0.27$ & $65.10/28 \approx 2.33$ \\
$\pi^- p$ CX & $1.051 \pm 0.022$ & $-0.49 \pm 0.41$ & $69.88/37 \approx 1.89$ \\
\hline
\end{tabular}
\end{center}
\vspace{0.5cm}
\end{table}

\begin{figure}
\begin{center}
\includegraphics [width=15.5cm] {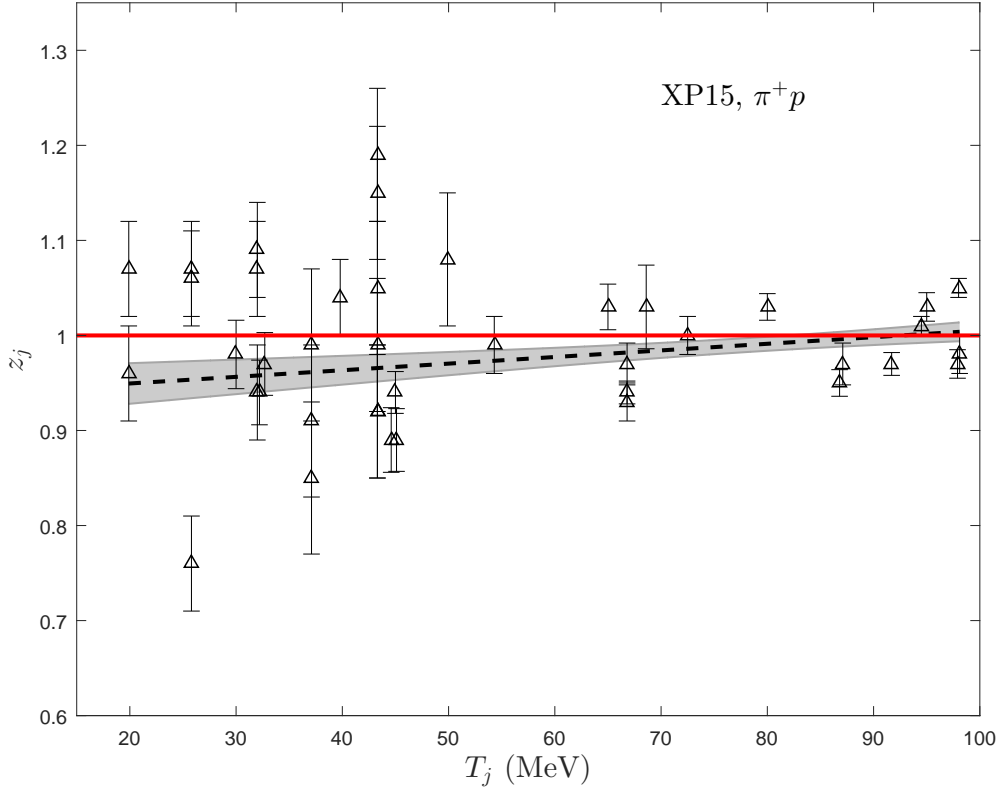}
\caption{\label{fig:PIPELXP15}Plot of the scale factors $z_j$ of Eq.~(\ref{eq:EQ004}) corresponding to the XP15 solution \cite{XP15}. $T_j$ denotes the pion laboratory kinetic energy of the $j$-th dataset. The data 
correspond to $\pi^+ p$ DCSs only. The datasets which were practically floated (the SAID group assign a normalisation uncertainty of $100~\%$ to datasets of suspicious absolute normalisation) have not been used. The dashed 
straight line represents the optimal, weighted linear fit to the data shown (see Table \ref{tab:XP15Parameters}), whereas the shaded band represents $1 \sigma$ uncertainties. The red line is the ideal outcome of the 
optimisation.}
\vspace{0.25cm}
\end{center}
\end{figure}

\begin{figure}
\begin{center}
\includegraphics [width=15.5cm] {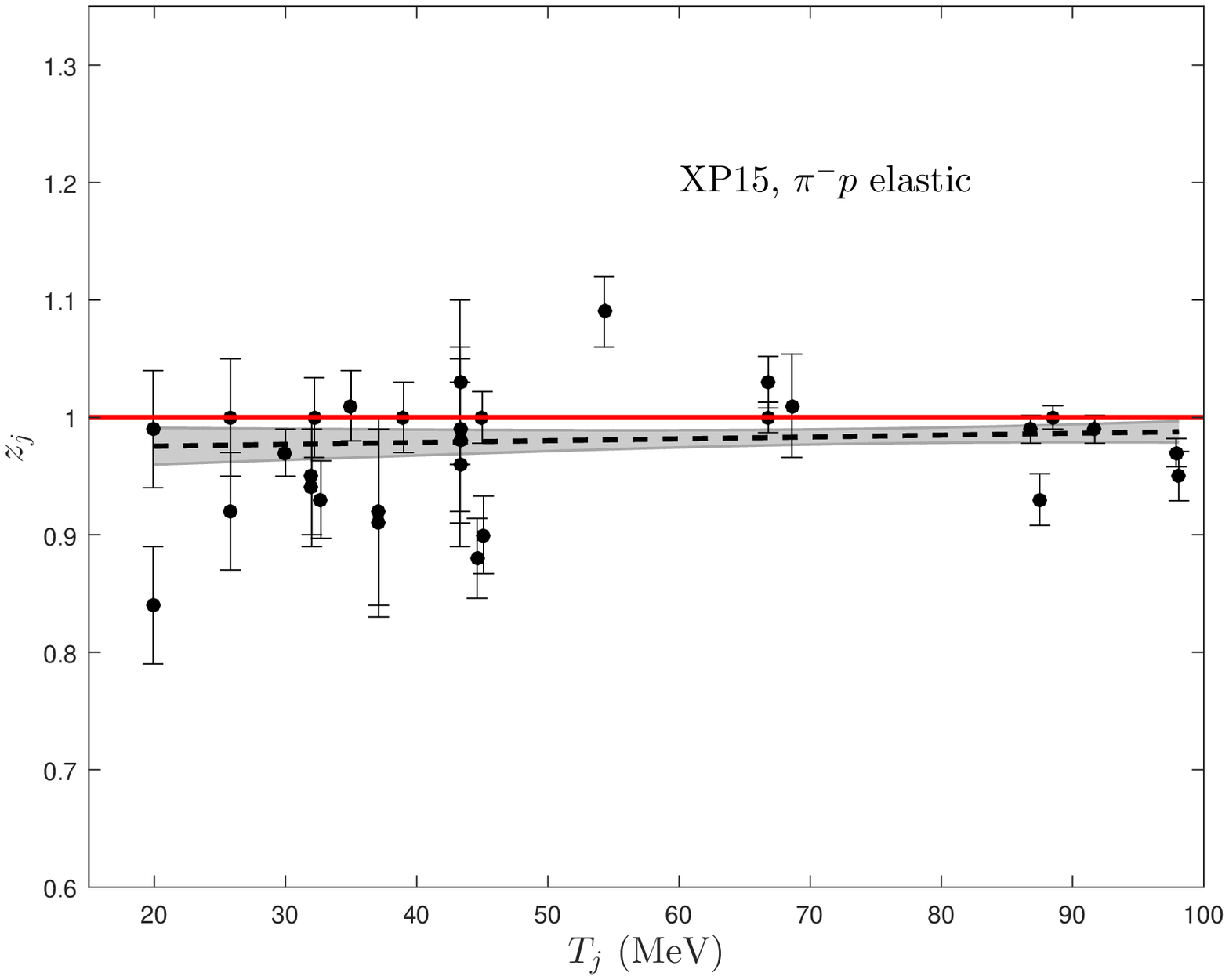}
\caption{\label{fig:PIMELXP15}The equivalent of Fig.~\ref{fig:PIPELXP15} for the SAID $\pi^- p$ ES DB.}
\vspace{0.25cm}
\end{center}
\end{figure}

\begin{figure}
\begin{center}
\includegraphics [width=15.5cm] {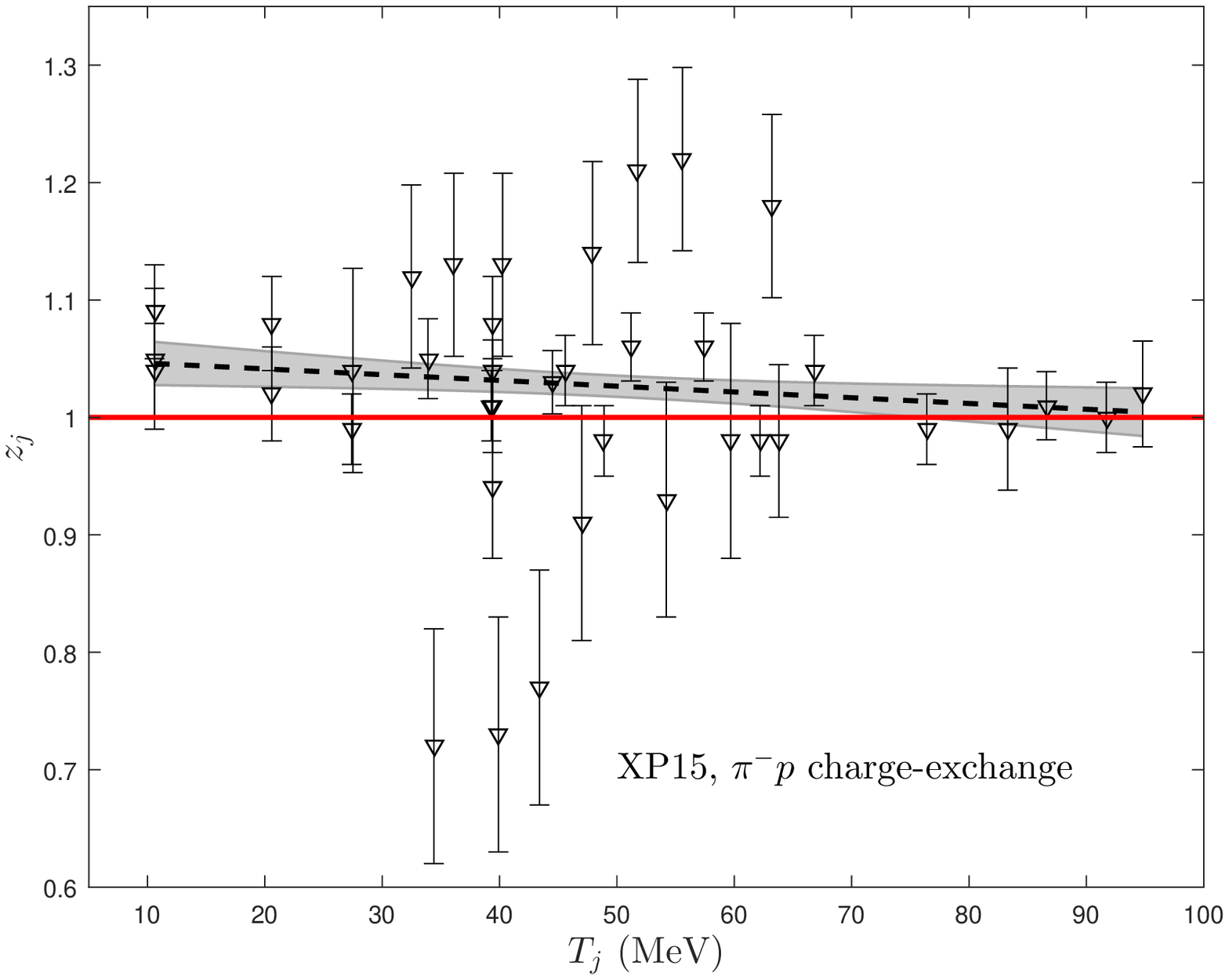}
\caption{\label{fig:PIMCXXP15}The equivalent of Fig.~\ref{fig:PIPELXP15} for the SAID $\pi^- p$ CX DB.}
\vspace{0.25cm}
\end{center}
\end{figure}

However, the SAID $\pi N$ DB comprises three distinct parts, i.e., corresponding to the three low-energy $\pi N$ reactions. Had their fit been truly satisfactory, a similar behaviour of the scale factors $z_j$ (in 
terms of their energy dependence), which is seen in Fig.~\ref{fig:ALLXP15}, should also have been observed in any arbitrary subset of their DB, complying with the basic principles of the Sampling Theory (adequate population, 
representative sampling). The fitted values of the scale factor $z$, relating to the description of the SAID low-energy $\pi N$ DBs with the XP15 solution, are shown (separately for the three reactions) in 
Figs.~\ref{fig:PIPELXP15}-\ref{fig:PIMCXXP15}. Had the three conditions (a)-(c), which are mentioned at the end of the previous section, been fulfilled in the SAID analysis at all energies, the fitted values of the scale 
factor $z$ of Figs.~\ref{fig:PIPELXP15}-\ref{fig:PIMCXXP15} should have come out independent of the beam energy and should have been centred on $1$ (as the case was for the results of the global fit of Fig.~\ref{fig:ALLXP15}). 
However, the bulk of the data for $T \leq 100$ MeV (represented by the shaded bands in Figs.~\ref{fig:PIPELXP15}-\ref{fig:PIMCXXP15}) seem to be either underestimated by the XP15 solution (i.e., the measurements of the 
$\pi^- p$ CX reaction) or overestimated by it (i.e., the ES measurements, the effects for the $\pi^+ p$ reaction being more pronounced), see also Table \ref{tab:XP15Parameters}. One notices that the difference to the 
statistical expectation decreases with increasing beam energy, vanishing in the vicinity of $T=100$ MeV. Such behaviour is in general agreement with the conclusions of Refs.~\cite{Matsinos2017a,Gibbs1995,Matsinos1997} for 
an energy-dependent isospin-breaking effect.

From Figs.~\ref{fig:PIPELXP15}-\ref{fig:PIMCXXP15} and from Table \ref{tab:XP15Parameters}, one may conclude that the XP15 solution does not describe sufficiently well the bulk of the low-energy measurements. Evidently, the 
XP15 solution at low energy represents a fictitious, average $\pi N$ process, one which does not adequately capture the dynamics of the three low-energy $\pi N$ reactions in sufficient detail.

\begin{figure}
\begin{center}
\includegraphics [width=15.5cm] {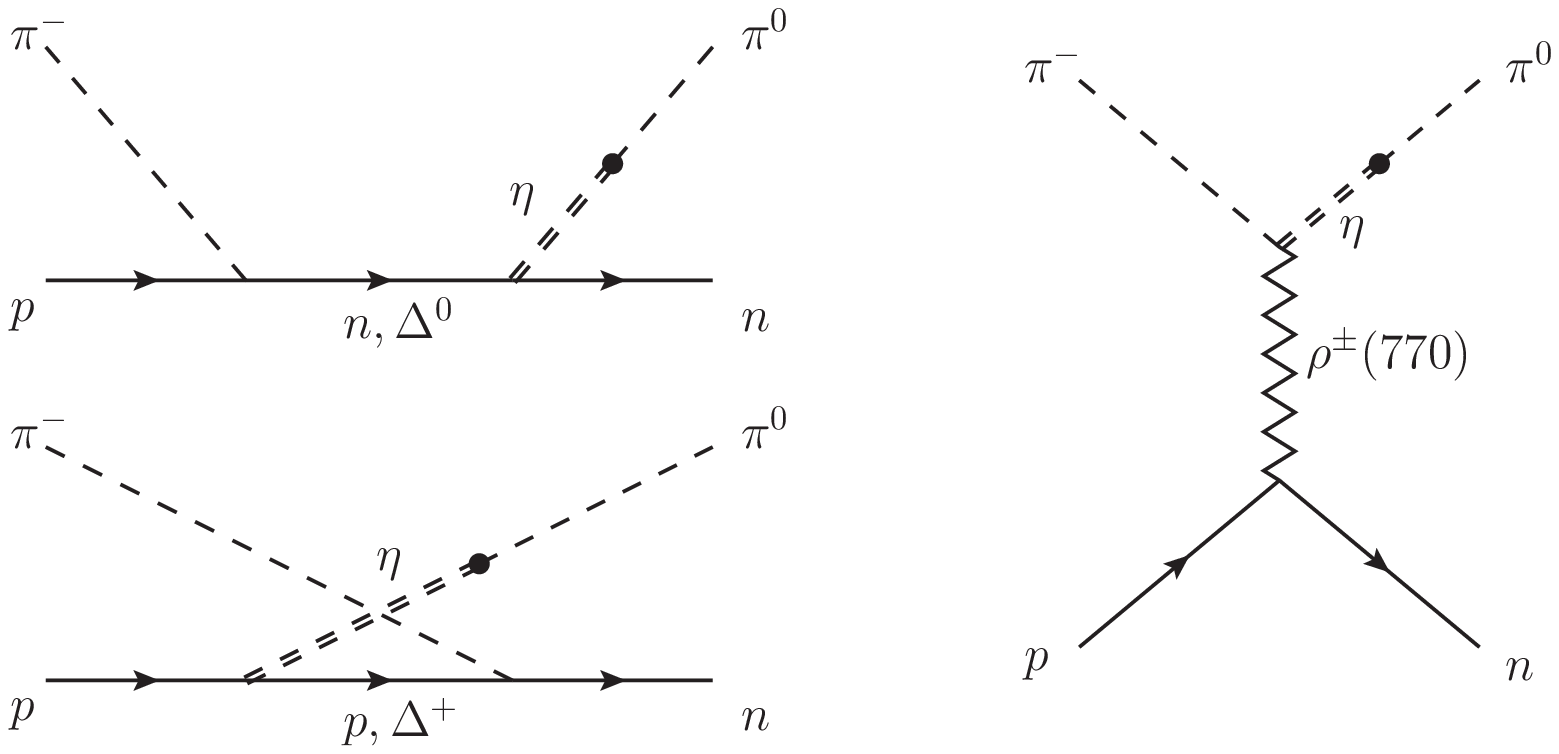}
\caption{\label{fig:IsospinBreakingEtaPi0}Feynman diagrams involving the $\pi^0 - \eta$ admixture, a mechanism for the violation of the isospin invariance in the hadronic part of the $\pi N$ interaction in case of the $\pi^- p$ 
CX reaction \cite{Cutkosky1979}.}
\vspace{0.25cm}
\end{center}
\end{figure}

In my judgment, only the ES data can jointly be submitted to an optimisation. There is no indication of violation of the isospin invariance in these two reactions. One of the possibilities in which models, based on hadronic 
exchanges, could accommodate a departure from the isospin invariance would have to involve $\rho^0-\omega$ mixing Feynman diagrams; however, the impact of this mechanism is suppressed at low energy, see Ref.~\cite{Matsinos2018} 
and the works cited therein. Experience shows that any optimisation, involving the $\pi^- p$ CX DB, departs from the expected statistical norms, resulting in a sizeable increase in $\chi^2_{\rm min}$ and in a pronounced 
energy dependence of the scale factors $z_j$. It ought to be reminded that the $\pi^0 - \eta$ mixing mechanism was proposed over four decades ago \cite{Cutkosky1979} as a potential source of isospin-breaking effects in the 
$\pi^- p$ CX reaction. It may be argued that, as the mechanism affects nearly all Feynman diagrams of the ETH model (see Fig.~\ref{fig:IsospinBreakingEtaPi0}), significant bias in the results might be expected whenever the 
data of the $\pi^- p$ CX reaction are included in the input DB. Although the coupling of the $\eta$ meson (compared with that of the pion) to the nucleon could be stronger, the range of the published $g_{\eta N N}$ values 
is fairly large. An average value $g_{\eta N N}=3.71(71)$ may be obtained from the entries of Table 2 of Ref.~\cite{Singh2019}, translating into the pseudovector coupling $f^2_{\eta N N}=0.093(36)$ or $f_{\eta N N}=0.306(58)$; 
in comparison, $f_c^2 \coloneqq f^2_{\pi^\pm p n} = 0.07629^{+0.00065}_{-0.00062}$ \cite{Matsinos2019}. It thus seems that the $\pi^0 - \eta$ transition could have a significant impact on the $\pi N$ dynamics, at least at 
low energy.

The conclusion seems to be inevitable. Even in analyses, which assume that the isospin invariance holds and pursue global fits to the data (as the case is for the SAID solutions), the isospin-breaking effects, albeit 
somewhat hidden, manifest themselves as a systematic bias in the output of the optimisation, and can be uncovered after the results of the optimisation for the three low-energy $\pi N$ reactions are \emph{separately} 
analysed.

\subsection{\label{sec:InadequateModelling}Can the low-energy $\pi N$ enigma be explained in terms of inadequate modelling of the hadronic part of the $\pi N$ interaction?}

On account of several reasons, inadequate modelling of the hadronic part of the $\pi N$ interaction is the most unlikely of the explanations, which could be put forward in order to resolve the low-energy $\pi N$ enigma. To 
start with, the approach of Ref.~\cite{Gibbs1995} is general and transparent, and the results, obtained on the basis of the hadronic potentials in that study, seem to be consistent among themselves. The approach of 
Refs.~\cite{Fettes1997,Matsinos2022a} is devoid of theoretical constraints, save for the expected behaviour of the $s$- and $p$-wave $K$-matrix elements at low energy. The optimal values of the coefficients of the polynomials, 
used in the parameterisation of these quantities, are obtained from optimisations which terminate successfully, posing no convergence problems whatsoever.

The ETH model contains Feynman diagrams of all well-established hadrons as intermediate states. Although the completeness of such a model can never be ascertained (in fact, one might also argue that all such models are bound 
to be `incomplete'), any residual contributions (e.g., from higher baryon resonances) cannot be significant (due to the largeness of their rest masses). In addition, the results obtained with the ETH model may be (and, in 
most cases, have been) verified by similar analyses performed using the approach of Refs.~\cite{Fettes1997,Matsinos2022a} (which, in fact, was the main reason that that approach had been put forward in the first place over 
two and a half decades ago).

The modelling of the hadronic part of the $\pi N$ interaction is different in Refs.~\cite{Matsinos2017a,Gibbs1995,Matsinos1997}: it is based on hadronic potentials in the work of Ref.~\cite{Gibbs1995}, whereas the two other 
works featured the ETH model. The results obtained with the ETH model are routinely compared with those extracted when using the polynomial parameterisation of the $s$- and $p$-wave $K$-matrix elements \cite{Fettes1997,Matsinos2022a}. 
The inclusion of the EM effects in Ref.~\cite{Gibbs1995} was based on algorithms developed by Gibbs and collaborators, whereas Refs.~\cite{Fettes1997,Matsinos1997} had used the NORDITA corrections. All analyses performed 
within the ETH $\pi N$ project after 2000 have used the EM corrections developed at the University of Zurich \cite{Oades2007,Gashi2001a,Gashi2001b}. The low-energy $\pi N$ DBs, which all these efforts rested upon, are 
different, enhanced as time went by. As a result, it appears to be remarkable that the analyses of Refs.~\cite{Gibbs1995,Matsinos1997} had come up with the same statistical significance of the discrepancy in the $\pi N$ 
interaction at low energy, equivalent to a $4 \sigma$ effect in the normal distribution; the statistical significance of the effect improved over the years, though the level of the discrepancy itself remained virtually 
constant \cite{Matsinos2017a}.

Regarding the studies \cite{Gibbs1995,Matsinos1997}, the occasionally-voiced criticism (mainly by $\chi$PT-inspired researchers) is that such models are inconsistent in their treatment of the EM and hadronic effects. It is 
true that within the ETH $\pi N$ project, the EM corrections to the $\pi N$ phase shifts and partial-wave amplitudes must be imported from external sources; I am not sure that the criticism is entirely justified in case of 
the work of Gibbs and collaborators \cite{Gibbs1995}. In any case, the same argument can be advanced against the plethora of works on the $\pi N$ interaction, which had been carried out within the $\chi$PT framework, 
but had used the SAID $\pi N$ phase shifts as input. In addition, though $\chi$PT is habitually presented in studies in ways which overstress its qualities, it nonetheless suffers from the same problems which plague all other 
effective-field methods. The evaluations of physical quantities are complete at one order, if they contain the contributions from all relevant physical effects at that order; in this sense, $\chi$PT provides a systematic 
basis for the treatment of the various physical effects. However, until the calculations are carried out at the next order, it remains unknown what changes they might bring. Furthermore, the convergence cannot easily be 
ascertained; it is usually assumed when two evaluations (of a physical quantity) at successive orders differ by an amount which is smaller than a user-defined threshold, though there is little assurance that the results of 
such evaluations should necessarily form a monotonic sequence. Even if a monotonic behaviour is expected in a problem (e.g., due to the smallness of the expansion parameter and the expected behaviour of the contributions 
from the physical effects entering that problem) and the convergence criterion has been met, the importance of the residual effects, i.e., those corresponding to all higher orders (above the one at which the evaluation was 
last carried out), can hardly be assessed.

\section{\label{sec:Explanations}How could the low-energy $\pi N$ enigma be explained?}

\subsection{\label{sec:Experimental}Experimental shortcomings}

The first attempt to provide an explanation for the low-energy $\pi N$ enigma involves an obvious effect, namely the faulty assessment of the absolute normalisation of the low-energy $\pi N$ datasets. Is such a possibility 
tenable? After all, there have been problems with $\pi N$ datasets at low energy over the years, involving bizarre angular distributions of the DCS, but also erroneous absolute normalisation. However, what is of relevance 
in this section is a \emph{systematic} misapplication of corrections to the raw experimental data~\footnote{For the sake of example, such a situation could arise if a part of the final-state pions evaded detection or if the 
flux of the incident beam were overestimated in the $\pi N$ experiments at low energy. In both cases, the DCS would be systematically underestimated.}, i.e., effects which impact on the \emph{bulk} of the measurements, not 
sporadic experimental shortcomings. Given that the experiments were conducted at different places and at times spanning nearly four decades, and that they involved various experimental groups, beamlines, targets, and 
detectors, the possibility of systematic experimental flaws, affecting the bulk of the measurements, appears to be doubtful. Having said that, let me elaborate further on two common (and related) concerns regarding the 
outcome of the $\pi N$ experimentation at low energy.

One concern is about the absolute normalisation of some low-energy experiments. One feels being at a loss to come up with an explanation for the significant effects in the rescaling of some of the datasets, which the PSAs 
of the $\pi N$ data, performed within both the SAID and the ETH $\pi N$ projects, occasionally reveal. (Those who wish to object should take a better look at the fluctuation of the plotted data in Fig.~\ref{fig:ALLXP15}.) 
The significant departure of the fitted values of the scale factor $z$ from $1$ for some of the datasets may be due to at least one of the following reasons.
\begin{itemize}
\item The energy of the incoming beam had not been what the experimental group anticipated~\footnote{For the sake of example, corrections were applied in the early 1990s to the energy calibration of the M11 channel at 
TRIUMF. I am not aware of formal corrections at other meson factories.}.
\item The effects of the contamination of the incoming beam had been underestimated in the experiment.
\item The normalisation uncertainty in the experiment had been underestimated.
\item The determination of the absolute normalisation in the experiment had been erroneous.
\end{itemize}

Another concern relates to the smallness of the normalisation uncertainty reported in some experiments. Arguments have been presented \cite{Matsinos2006}, to substantiate the point of view that it is very likely that the 
published uncertainties in the $\pi N$ experiments at low energy had been underestimated \emph{on average}. For instance, for nearly half of the DCS datasets (with known normalisation uncertainty) in the ES DB, normalisation 
uncertainties below $3~\%$ had been reported. The smallest normalisation uncertainty in that DB is equal to a mere $1.2~\%$, which (based on the experience gained after three decades of relevant experimentation) borders on 
the impossible (at present). While pondering over these issues, one cannot help thinking that it would make sense to disregard all claims of such exaggerated precision, and to assign to all relevant datasets a reasonable 
normalisation uncertainty, say, $3~\%$ (though this value might be optimistic too). However, such an approach would seem to be arbitrary and would undoubtedly (and, to an extent, justifiably) provide ground for criticism. 
Such revisions ought to be instigated by the experimental groups which had been responsible for the measurements, not by analysts. Regrettably, it is unrealistic to expect that such effects could be re-examined, in particular 
with a critical eye, given the time which elapsed since most of the $\pi N$ experiments at low energy were conducted, the unfortunate loss of relevant information, and the shifting interests of the experimental groups towards 
other research domains as Pion Physics gradually phased out during the recent years.

\begin{figure}
\begin{center}
\includegraphics [width=15.5cm] {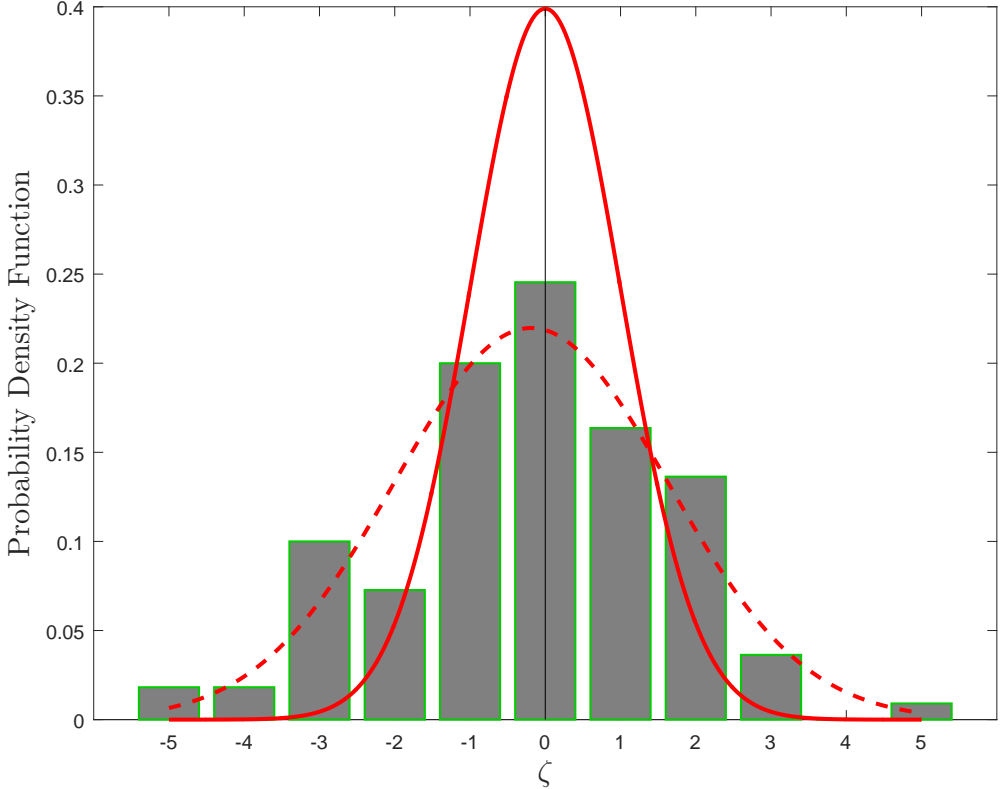}
\caption{\label{fig:PDF}The probability density function (PDF) of the normalised residuals $\zeta_j$, corresponding to the scale factors $z_j$ (and the associated normalisation uncertainties $\delta z_j$) in the SAID 
low-energy $\pi N$ DB. The quantities $\zeta$ are expected to follow the $N(0,1)$ distribution, represented in the plot by the continuous red curve. The dashed red curve is the PDF of the distribution $N(\hat{\mu},\hat{\sigma}^2)$, 
where $\hat{\mu} \approx -0.19$ and $\hat{\sigma} \approx 1.82$, corresponding to the SAID output. This plot strengthens the supposition that the normalisation effects had been underestimated \emph{on average} in the $\pi N$ 
experiments at low energy.}
\vspace{0.25cm}
\end{center}
\end{figure}

The fluctuation, which is present in Fig.~\ref{fig:ALLXP15}, attests to the seriousness of these concerns; the result of the linear fit $\chi^2 \approx 355.45$ for $108$ degrees of freedom indicates a poor description of 
the data, suggestive of erroneous absolute normalisation of and/or underestimated normalisation uncertainty in many low-energy $\pi N$ datasets. On the other hand, the plot also indicates that some of the problematic 
datasets have overestimated absolute normalisation, whereas a similar number have underestimated one. On the whole, the two categories balance one another (which, in fact, is favoured due to the use of the Arndt-Roper 
minimisation function, as opposed to more robust procedures, in the optimisation), and the distribution of the fitted values of the scale factor $z$ comes out nearly centred on $1$. Let me next examine the distribution of 
the normalised residuals of the scale factor $z$, defined as: $\zeta_j \coloneqq (z_j - 1) / \delta z_j$. Ideally, $\zeta$ should be normally distributed, i.e., it should follow the $N(\mu=0,\sigma^2=1)$ distribution. The 
average $\zeta$ value, obtained from the $110$ low-energy $\pi N$ datasets in the SAID DB, comes out equal to $\hat{\mu}=-0.19(17)$, suggesting no significant departure from a symmetrical distribution (about $0$). On the 
contrary, for the square root of the unbiased variance, one obtains $\hat{\sigma} \approx 1.82$, suggesting a sizeable departure of the $\zeta$ distribution from the $N(0,1)$ distribution, see Fig.~\ref{fig:PDF}. From these 
results, one can hypothesise that the large reduced $\chi^2$ value for the description of the data in Fig.~\ref{fig:ALLXP15} is due to smaller $\delta z_j$ values having been used in the definition of the normalised residuals 
$\zeta_j$. Absorbing the fluctuation in Fig.~\ref{fig:ALLXP15} into a redefinition of the normalisation uncertainty $\delta z$ favours the scenario that that quantity had been underestimated \emph{on average} in the 
low-energy $\pi N$ experimentation, by possibly as much as $45~\%$ (i.e., by a factor of nearly $2$). Last but not least, the three low-energy $\pi N$ reactions are not equally affected by the systematic underestimation of 
the normalisation uncertainty of the experiments; Table \ref{tab:XP15Parameters} leaves no doubt that the imprint of the effect on the $\pi^+ p$ experiments is deeper.

To summarise, a systematic bias in the absolute normalisation of the low-energy $\pi N$ data, one which would impact on the bulk of the available measurements, however unlikely, cannot be altogether removed from 
consideration. The sizeable fluctuation in Fig.~\ref{fig:ALLXP15} (or, equivalently, the presence of substantial tails in the distribution of the normalised residuals $\zeta$ in Fig.~\ref{fig:PDF}) indicates that it is 
very likely that the normalisation uncertainty had been underestimated \emph{on average} in the $\pi N$ experiments at low energy. Be that as it may, `outsiders' cannot resolve such issues; any experimental shortcomings ought 
to be addressed (and resolved) by the members of the original experimental groups, a prospect which seems to be remote at present.

\subsection{\label{sec:ResidualEM}Residual EM contributions}

The completeness of the EM corrections in the $\pi N$ interaction is an important issue which must be addressed with the seriousness it deserves. The way I understand this matter in the context of a Universe $\mathscr{A}$, 
in which only hadronic and EM phenomena are of relevance, may be summarised in one sentence. If complete, an EM correction to a value of a physical quantity, obtained in Universe $\mathscr{A}$, will translate it into the 
corresponding result in Universe $\mathscr{B}$, which is devoid of all EM effects. If my understanding/interpretation is correct, then one may pose an inevitable question: what happens to the particle `proton' itself when 
the EM interaction is switched off? Without doubt, the rest mass of the proton receives EM contributions, which ought to be deducted in Universe $\mathscr{B}$. Therefore, the particle `proton' of Universe $\mathscr{A}$ will 
have another mass in Universe $\mathscr{B}$ (and, of course, will be neutral). The same applies to all other (charged or composite) particles, namely to the `neutron' and to the `pions'. If the strong interaction treats the 
members of isospin multiplets on an equal footing, then the hadronic masses of protons and neutrons must be identical. The same applies to the charged and neutral pions, which must share one hadronic mass. Therefore, the 
four rest masses of Universe $\mathscr{A}$ (proton, neutron, charged pion, and neutral pion) would reduce to two hadronic masses in Universe $\mathscr{B}$, namely the hadronic mass of the nucleon and that of the pion. The 
former ought to be smaller than the rest mass of the proton, whereas the latter ought to be smaller than (or equal to?) the rest mass of the neutral pion. Some suggest that the hadronic mass of the pion should be taken to 
be the rest mass of the neutral pion. That would be a breakthrough (as one hadronic mass in the $\pi N$ interaction would be fixed), but I believe that one can argue further, against this possibility, in that a neutral pion 
consists of $q \bar{q}$ pairs. Evidently, as we descended one level into the structure of matter, another question is posed, while the original one remains unanswered: what are the EM contributions to the `physical' masses 
of the quarks?

In Ref.~\cite{Matsinos2006}, a clear distinction is made between stage-1 and stage-2 EM corrections. The stage-1 EM corrections should provide estimates for the so-called trivial EM effects (direct Coulomb amplitude, 
effects of the extended charge distributions of the interacting hadrons, vacuum polarisation, and all relevant interferences) and, in the case of $\pi^- p$ scattering, for the external mass differences and the effects of 
the $\gamma n$ channel. The three studies on the EM corrections, which were carried out at the University of Zurich in the late 1990s and the early 2000s, aimed at the removal of the stage-1 EM effects in low-energy 
scattering \cite{Gashi2001a,Gashi2001b}, as well as at the $\pi N$ threshold \cite{Oades2007}, in a consistent manner. On the other hand, the stage-2 EM corrections should go one step further: in addition to Feynman 
diagrams with loops and internal photon lines, they should also take care of the effects relating to the use of the physical rest masses of the particles in the stage-1 EM corrections, instead of the hadronic ones.

Needless to emphasise that the available EM-correction schemes in the scattering region \cite{Gibbs1998,Tromborg1976,Tromborg1977,Tromborg1978,Gashi2001a,Gashi2001b} aim at the removal of the trivial effects relating to 
the EM interaction of the particles involved, but assume that no change is induced on the particles themselves as the result of the `switching-off' of the EM interaction. The underlying assumption in all aforementioned 
studies is that the hadronic masses of the various particles are the corresponding physical ones. At this point, one basic question calls for a satisfactory answer: are the EM corrections supposed to also remove the EM 
contributions to the rest masses? If the answer to this question is `yes', then all known EM-correction schemes are incomplete, and all analyses of the $\pi N$ data had been pleasant assignments, largely reminiscent of 
Tinguely's contraptions~\footnote{Jean Tinguely (1925-1991) was a Swiss sculptor, whose creations I always considered highly innovative and totally useless.}. If the answer to the aforementioned question is `yes', then the 
hadronic part of the $\pi N$ interaction is not sufficiently well known at present.

I believe that knowledge about the hadronic part of the $\pi N$ interaction will not be furthered by new experiments (even if they could be conducted somewhere), while the question of the EM corrections remains unresolved. 
Even for perfect data and no discrepancies whatsoever in the DB, one would still need to address the extraction of the important (hadronic) information from those perfect measurements. Important issues, which ought to be 
addressed in the reassessment of the EM effects in the $\pi N$ interaction, include the following.
\begin{itemize}
\item Clear definitions of what is meant by `EM part' and by `hadronic part' of the $\pi N$ interaction.
\item Suitable methodology to deal with the hadrons after they (or their constituents, i.e., the quarks and antiquarks) have been deprived of their EM features.
\item Suitable methodology to deal with the altered kinematics, after the interacting hadrons have been deprived of the EM contributions to their rest masses.
\end{itemize}

To summarise, an advancement of knowledge in Pion Physics could only be instigated by a theoretical breakthrough, in particular in relation to the reliable removal of the EM effects.

In the subsequent section, I will demonstrate that, due to the sizeably different results of the various EM-correction schemes (at present), even the extraction of the values of the $\pi N$ scattering lengths from the 
accurate PSI measurements at the $\pi N$ threshold is not as precise as it should have been. And, if the evaluation of the EM effects at the $\pi N$ threshold is controversial to the extent that the hadronic $\pi N$ 
scattering lengths are not sufficiently well known, there can be no doubt that the $\pi N$ phase shifts are even less `well-known'. In my opinion, prior to the development of advanced methods to extrapolate the $\pi N$ 
scattering amplitudes to the unphysical region, the theoretical innovation should be directed to the physical one.

\subsubsection{\label{sec:ResidualEMAtThreshold}How does the possibility of residual EM effects gain momentum in the light of the reported corrections at the $\pi N$ threshold?}

The first of the Deser formulae relates the strong shift $\epsilon_{1 s}$ in pionic hydrogen with the `untreated' (i.e., containing effects of EM origin) scattering length $a_{cc}$. Following the sign convention of Ref.~\cite{Schroeder2001},
\begin{equation} \label{eq:EQ006}
\epsilon_{1 s} \coloneqq E^{\rm EM}_{np \to 1s} - E^{\rm measured}_{np \to 1s} =- 4 \frac{ \lvert E_{1 s} \rvert}{r_B} a_{cc} = - \frac{2 \alpha^3 \mu^2}{\hbar c} a_{cc} \, \, \, ,
\end{equation}
where $E_{1 s}= - \alpha^2 \mu / 2$ is the (point-Coulomb) EM binding energy of the $1 s$ level and $r_B = \hbar c / (\alpha \mu)$ is the Bohr radius; $\alpha$ denotes the fine-structure constant and $\mu$ stands for the 
reduced mass of the $\pi^- p$ system.

The second of the Deser formulae, put into its current form by Trueman \cite{Trueman1961}, enables the extraction of the scattering length $a_{c0}$ from the total decay width $\Gamma_{1 s}$ in pionic hydrogen.
\begin{equation} \label{eq:EQ007}
\Gamma_{1 s}=8 q_0 \frac{\lvert E_{1 s} \rvert}{r_B \hbar c} \left( 1 + P^{-1} \right) a_{c0}^2 = 4 q_0 \frac{\alpha^3 \mu^2}{(\hbar c)^2} \left( 1 + P^{-1} \right) a_{c0}^2 \, \, \, ,
\end{equation}
where $q_0$ denotes the magnitude of the CM $3$-momentum of the outgoing $\pi^0$ (or neutron) at the $\pi^- p$ threshold and $P=1.546(9)$ \cite{Spuller1977} is the Panofsky ratio, the ratio between the CX and the 
radiative-capture partial decay widths: $P \coloneqq \Gamma_{1 s} (\pi^- p \to \pi^0 n) / \Gamma_{1 s} (\pi^- p \to \gamma n)$.

Corrections must be applied, to rid $a_{cc}$ and $a_{c0}$ of all effects of EM origin, and lead to estimates for the hadronic scattering lengths $\tilde{a}_{cc}$ and $\tilde{a}_{c0}$. The EM corrections are usually 
expressed in the form of two quantities, $\delta_\epsilon$ for $a_{cc}$ and $\delta_\Gamma$ for $a_{c0}$. The two hadronic scattering lengths are obtained from $a_{cc}$ and $a_{c0}$ according to the following definitions.
\begin{equation} \label{eq:EQ008}
\tilde{a}_{cc} = a_{cc} / (1 + \delta_\epsilon)
\end{equation}
\begin{equation} \label{eq:EQ009}
\tilde{a}_{c0} = a_{c0} / (1 + \delta_\Gamma)
\end{equation}

It is understood that the scattering lengths $a_{cc}$ and $a_{c0}$ in Eqs.~(\ref{eq:EQ008},\ref{eq:EQ009}) are associated with the original Deser formulae (\ref{eq:EQ006},\ref{eq:EQ007}). These formulae represent 
\emph{leading-order} (LO) evaluations of $\epsilon_{1 s}$ and $\Gamma_{1 s}$, namely determinations at $\mathcal{O}(\alpha^3)$. In several works, the quantities $a_{cc}$ and $a_{c0}$ of Eqs.~(\ref{eq:EQ006},\ref{eq:EQ007}) 
are therefore denoted as $a_{cc}^{\rm LO}$ and $a_{c0}^{\rm LO}$, as opposed to the quantities entering the upgraded forms of the Deser formulae, i.e., the expressions obtained at higher orders of $\alpha$. At present, 
only the next-to-leading-order (NLO) evaluations of $\epsilon_{1 s}$ and $\Gamma_{1 s}$ are available, i.e., the evaluations at $\mathcal{O}(\alpha^4)$. Some authors denote the scattering lengths, entering the NLO 
evaluations of $\epsilon_{1 s}$ and $\Gamma_{1 s}$, as $a_{cc}^{\rm NLO}$ and $a_{c0}^{\rm NLO}$. In this work, $a_{cc}$ and $a_{c0}$ will represent $a_{cc}^{\rm LO}$ and $a_{c0}^{\rm LO}$, respectively. I will introduce 
$\mathcal{A}_{cc}$ and $\mathcal{A}_{c0}$ later on, when referring to $a_{cc}^{\rm NLO}$ and $a_{c0}^{\rm NLO}$, respectively.

Several schemes of removal of the EM effects from the measurements were developed after the first experimental results at the $\pi N$ threshold became available; some of these schemes aim at the removal of the trivial EM 
effects, i.e., of those comprising stage-1 EM corrections. The models of Sections 5.2.1.1 and 5.2.1.2 are supposed to remove only these effects. On the contrary, works carried out within the $\chi$PT framework also 
attempt the removal of effects which are associated with the mass difference of the two light quarks ($u$ and $d$). The models of Sections 5.2.1.3-5.2.1.6 belong to this category. When comparing the results of the various 
schemes, one ought to bear in mind the distinction between these two categories of corrections.

\emph{5.2.1.1 Potential models for the removal of the EM effects}

In their assessment of the EM effects at the $\pi N$ threshold, Refs.~\cite{Sigg1996b,Oades2007} made use of suitable hadronic potentials~\footnote{Regrettably, I failed to mention the 1987 paper of Kaufmann and Gibbs 
\cite{Kaufmann1987} in the first version of this preprint. Comments on that work may be found in Refs.~\cite{Sigg1996b,Oades2007}.}.

An estimate for $\delta_\epsilon$ was obtained in Ref.~\cite{Sigg1996b} by means of a two-channel calculation, along with the phenomenological inclusion of the effects of the $\gamma n$ channel: $\delta_\epsilon=-2.1(5) \cdot 10^{-2}$. 
Using a genuine three-channel calculation, Ref.~\cite{Oades2007} obtained an incompatible result: $\delta_\epsilon=0.67(67) \cdot 10^{-2}$. The difference in the $\delta_\epsilon$ results between the two studies is due to 
the different treatment of the effects of the $\gamma n$ channel.

In Ref.~\cite{Sigg1996b}, the estimate for $\delta_\Gamma$ of $-1.3(5) \cdot 10^{-2}$ had been extracted. On the other hand, Ref.~\cite{Oades2007} obtained: $\delta_\Gamma=-1.66(33) \cdot 10^{-2}$. Therefore, the two 
$\Gamma_{1 s}$ corrections \cite{Sigg1996b,Oades2007} agree within the uncertainties; evidently, the correction $\delta_\Gamma$ is less sensitive to the way by which the effects of the $\gamma n$ channel are included in the 
calculation.

\emph{5.2.1.2 The model of Ericson and collaborators \cite{Ericson2004}}

In 2004, Ericson and collaborators \cite{Ericson2004} followed a non-relativistic approach using Coulomb wavefunctions, with a short-range strong interaction and extended charge distributions, and treated four sources of 
EM corrections. The first two take account of effects relating to the vacuum polarisation and the extended charge distributions of the interacting particles. The remaining two corrections relate to renormalisation and gauge 
effects: the former takes account of the continuity and smoothness of the wavefunction on the spherical boundary which delimits the application of the strong interaction, whereas the latter correction is due to the 
adjustment of the energy level in such a way that it corresponds to the scattering off an extended charge Coulomb potential close to the origin.

The estimates of Ref.~\cite{Ericson2004} for the corrections $\delta_\epsilon$ and $\delta_\Gamma$ may be found in their Table 1: $\delta_\epsilon = -0.62(29) \cdot 10^{-2}$ and $\delta_\Gamma = 1.02(23) \cdot 10^{-2}$. The 
discrepancy in $\delta_\Gamma$ between the results, obtained with the potential models of Section 5.2.1.1 and with the model of this part, is noticeable. The correction $\delta_\epsilon$ of Ref.~\cite{Ericson2004} lies 
in-between the results of the two potential models of the previous section, slightly closer to the estimate of Ref.~\cite{Oades2007}.

The subject of the corrections $\delta_\epsilon$ and $\delta_\Gamma$ within the model of Ericson and collaborators \cite{Ericson2004} was recently revisited \cite{Matsinos2022b}, using the results of fits to a subset of the 
currently-available low-energy $\pi N$ measurements, rather than importing information from the outdated Karlsruhe analyses \cite{Hoehler1983} (which Ref.~\cite{Ericson2004} had actually done). As Ref.~\cite{Matsinos2022b} 
has not been subjected to the peer-review process, I decided to quote the original $\delta_\epsilon$ and $\delta_\Gamma$ estimates \cite{Ericson2004} in Table \ref{tab:EMCorrections} and to use the same estimates for the 
purposes of Figs.~\ref{fig:Acc}-\ref{fig:b1}.

Comments on the approach of Ref.~\cite{Ericson2004} may be found in Section 4 of Ref.~\cite{Oades2007}; there is no point in repeating them here. It must be borne in mind that Ericson and collaborators \cite{Ericson2004} had 
also expressed criticism about the models of Section 5.2.1.1, on the basis of the inconsistency of the coupled-channel formalism with the low-energy expansion of the $K$-matrix elements, see also Ref.~\cite{Ericson2002}.

\emph{5.2.1.3 The Lyubovitskij-Rusetsky correction to $a_{cc}$ \cite{Lyub2000}}

I consider the 2000 paper of Lyubovitskij and Rusetsky \cite{Lyub2000} important for two reasons.
\begin{itemize}
\item The authors presented a calculation of the strong shift $\epsilon_{1 s}$ at $\mathcal{O}(\alpha^4)$; this is an essential upgrade of Eq.~(\ref{eq:EQ006}). Part of the effects, which ought to be taken care of by the 
EM corrections in case that Eq.~(\ref{eq:EQ006}) is used, are contained in the upgraded expression.
\item Their work constituted the first attempt to determine the isospin-breaking corrections, which are due to the mass difference of the two light quarks, within the $\chi$PT framework.
\end{itemize}

The relation at $\mathcal{O}(\alpha^4)$ between $\epsilon_{1 s}$ and the (untreated) $\pi^- p$ ES length (denoted as $\mathcal{A}$ in Ref.~\cite{Lyub2000}, $\mathcal{A}_{cc}$ in this work) reads as:
\begin{equation} \label{eq:EQ010}
\epsilon_{1 s} = - \frac{2 \alpha^3 \mu^2}{\hbar c} \mathcal{A}_{cc} \left( 1 + 2 \alpha \left( 1 - \ln \alpha \right) \frac{\mu \mathcal{A}_{cc}}{\hbar c} \right) \, \, \, ,
\end{equation}
which, after appending the effects of the vacuum-polarisation correction to the ground-state wavefunction $\varphi \approx 0.483 \cdot 10^{-2}$ of Ref.~\cite{Eiras2000} (these effects had not been included in 
Ref.~\cite{Lyub2000}), may be rewritten as
\begin{equation} \label{eq:EQ011}
\epsilon_{1 s} = - \frac{2 \alpha^3 \mu^2}{\hbar c} \mathcal{A}_{cc} \left( 1 + \varphi + 2 \alpha \left( 1 - \ln \alpha \right) \frac{\mu \mathcal{A}_{cc}}{\hbar c} \right) \, \, \, .
\end{equation}
Lyubovitskij and Rusetsky did not identify $\mathcal{A}_{cc}$ with $\tilde{a}_{cc}$. Additional corrections (denoted as $\epsilon$ in Ref.~\cite{Lyub2000}, $\Delta \mathcal{A}_{cc}$ in this work), to be understood as 
contributions originating from residual EM effects and from the mass difference of the two light quarks, were evaluated in Ref.~\cite{Lyub2000} at $\mathcal{O}(p^2)$ in $\chi$PT. The relation between the quantities 
$\mathcal{A}_{cc}$ and $\tilde{a}_{cc}$ was given in Ref.~\cite{Lyub2000} as:
\begin{equation*}
\tilde{a}_{cc} = \mathcal{A}_{cc} - \Delta \mathcal{A}_{cc} \, \, \, .
\end{equation*}
The correction $\Delta \mathcal{A}_{cc}$ depends on three low-energy constants (LECs) $c_1$, $f_1$, and $f_2$, one of which ($f_1$) is poorly known. According to Ref.~\cite{Lyub2000}:
\begin{equation} \label{eq:EQ012}
\Delta \mathcal{A}_{cc} = \frac{m_p \hbar c}{2 (m_p+m_c)} \left( \frac{2 (m_c^2-m_0^2)}{\pi F_\pi^2} c_1 - \alpha (4 f_1 + f_2) \right) \, \, \, ,
\end{equation}
where $m_0$, $m_c$, and $m_p$ are the rest masses of the neutral pion, of the charged pion, and of the proton, respectively; $F_\pi=92.07(85)$ MeV is the pion-decay constant, see Eqs.~(71.13,71.14) in the 
chapter `Leptonic Decays of Charged Pseudoscalar Mesons' of Ref.~\cite{PDG2020} (their $f_\pi$ is equal to $\sqrt{2} F_\pi$).

\begin{itemize}
\item For $c_1$, Lyubovitskij and Rusetsky used a result from the Karlsruhe programme of the mid 1980s \cite{Hoehler1983}, privately communicated to the authors: that value was equal to $-0.925$ GeV$^{-1}$ (no uncertainty 
was quoted in Ref.~\cite{Lyub2000}). Also using input from the same source, Gasser and collaborators \cite{Gasser2002} came up (in 2002) with $c_1 = -0.93(7)$ GeV$^{-1}$. In the same year, Lyubovitskij and collaborators 
\cite{Lyub2002} imported (from a work of 2001) a different $c_1$ value, namely $c_1=-1.2(1)$ GeV$^{-1}$. More recent works \cite{Baru2011a,Baru2011b} recommend: $c_1 = -1.0(3)$ GeV$^{-1}$.
\item Regarding the LEC $f_2$, Ref.~\cite{Lyub2000} used $f_2 = - 0.97(38)$ GeV$^{-1}$, which is the recommended value in Ref.~\cite{Gasser2002}.
\item As aforementioned, the LEC $f_1$ is poorly known. To obtain an estimate for the correction $\delta_\epsilon$, Lyubovitskij and Rusetsky assumed in Ref.~\cite{Lyub2000} that $\lvert f_1 \rvert \leq \lvert f_2 \rvert$. 
However, Lyubovitskij and collaborators \cite{Lyub2002} arrived in 2002 at a mismatching result for the ratio $f_1/f_2$, namely $2.24(26)$. The authors favoured $f_1=-2.29(19)$ GeV$^{-1}$, which does not seem to be in 
line with the other `expectations' for this LEC. Gasser and collaborators \cite{Gasser2002} mention their ``order of magnitude'' estimate for $\lvert f_1 \rvert$ at about $1.4$ GeV$^{-1}$.
\end{itemize}
Be that as it may, Ref.~\cite{Lyub2000} reported a `large' negative correction: $\delta_\epsilon=(-4.8 \pm 2.0) \cdot 10^{-2}$. I set out to re-evaluate the correction $\delta_\epsilon$ in the Lyubovitskij-Rusetsky scheme, 
using Eq.~(\ref{eq:EQ011}), rather than Eq.~(\ref{eq:EQ010}) which the authors had used. As Ref.~\cite{Lyub2000} mentions no uncertainty in the LEC $c_1$, I first assumed that $c_1$ was not varied in their analysis. However, 
the resulting uncertainty of $\delta_\epsilon$ turned out to be nearly a factor of $2$ smaller than the one quoted in Ref.~\cite{Lyub2000}. Therefore, I concluded that also $c_1$ was varied in Ref.~\cite{Lyub2000} and 
proceeded by changing the assigned $c_1$ uncertainty, until the final result for $\delta_\epsilon$ matched the reported $\delta_\epsilon$ uncertainty of Ref.~\cite{Lyub2000}. My conclusion is that Lyubovitskij and Rusetsky 
had most likely used a $\delta c_1$ value between $0.2$ and $0.3$ GeV$^{-1}$ in their work. In any case, the $\delta c_1$ value of $0.3$ GeV$^{-1}$, also recommended in Refs.~\cite{Baru2011a,Baru2011b}, appears to be 
reasonable and conservative. For the needs of Table \ref{tab:EMCorrections}, I obtain the correction $\delta_\epsilon$ using Eq.~(\ref{eq:EQ011}) with the $\varphi$ value of Ref.~\cite{Eiras2000} and $\delta c_1 = 0.3$ 
GeV$^{-1}$. The other two LECs are varied according to Ref.~\cite{Lyub2000}.

As the models of Sections 5.2.1.1 and 5.2.1.2 do not contain any stage-2 EM corrections, the comparison of their $\delta_\epsilon$ values with the result of this part makes little sense. On the other hand, one could pose the 
question whether a comparison could be meaningful if $\tilde{a}_{cc}$ were identified with the scattering length $\mathcal{A}_{cc}$, obtained from Eq.~(\ref{eq:EQ011}). There is no doubt that some of the effects, which are 
treated by the models of Sections 5.2.1.1 and 5.2.1.2, are contained in $\Delta \mathcal{A}_{cc}$ of Eq.~(\ref{eq:EQ012}). Unfortunately, it is not clear to me how to disentangle the EM contributions and those relating to 
the $m_u \neq m_d$ effects in Eq.~(\ref{eq:EQ012}). Therefore, it seems that there is no assurance that a comparison of the results of this part with those obtained with the models of Sections 5.2.1.1 and 5.2.1.2 is 
meaningful. Nevertheless, I will also obtain a $\delta_\epsilon$ value corresponding to the case that $\mathcal{A}_{cc}$ of Eq.~(\ref{eq:EQ011}) is identified with $\tilde{a}_{cc}$. This intermediate result could be helpful 
later on, in assessing the importance of the isospin-breaking effects at $\mathcal{O}(p^2)$.

In 2002, Lyubovitskij and collaborators \cite{Lyub2002} provided an update of $\delta_\epsilon$, claiming improved knowledge of $f_1$ after deploying their ``perturbative chiral quark model;'' the new value was 
$\delta_\epsilon=-2.8 \cdot 10^{-2}$, quoted in Ref.~\cite{Lyub2002} without an uncertainty. Evidently, the updated value of Ref.~\cite{Lyub2002} is not incompatible with the 1996 result extracted with the potential model 
of Ref.~\cite{Sigg1996b}.

\emph{5.2.1.4 Isospin-breaking corrections evaluated at $\mathcal{O}(p^3)$ in $\chi$PT \cite{Gasser2002}}

An even larger (and more uncertain) correction $\delta_\epsilon$ was extracted in 2002 \cite{Gasser2002} within a calculation at NLO ($\mathcal{O}(p^3)$) in isospin breaking and in the low-energy expansion: 
$(-7.2 \pm 2.9) \cdot 10^{-2}$.

\emph{5.2.1.5 Leading-order correction $\delta_\Gamma$ in $\chi$PT \cite{Zemp2004}}

The LO correction to $a_{c0}$, obtained in 2004 within the $\chi$PT framework \cite{Zemp2004}, was found small: $\delta_\Gamma=0.6(2) \cdot 10^{-2}$, see Eq.~(5.26) therein. One notices that, within the $\chi$PT framework, 
the correction $\delta_\Gamma$ is more precisely known than $\delta_\epsilon$. This is due to the fact that the LECs $c_1$ and $f_1$ do not enter the determination of $\delta_\Gamma$, e.g., see Eq.~(\ref{eq:EQ014_5}).

\emph{5.2.1.6 The corrections developed by the Bonn-J{\"u}lich group}

Another correction scheme for the pionic-hydrogen measurements, along the lines of those detailed in the last three sections, was developed between 2005 and 2011, see Refs.~\cite{Baru2011a,Baru2011b} and the relevant papers 
therein. In addition, corrections for the strong shift of the $1 s$ state in pionic deuterium were advanced.

Regarding the strong shift $\epsilon_{1 s}$ in pionic hydrogen, Ref.~\cite{Baru2011a} uses Eq.~(\ref{eq:EQ011}) to extract $\mathcal{A}_{cc}$, which the authors call $a_{\pi^- p}$ in their paper. They subsequently associate 
$\mathcal{A}_{cc}$ with the difference $b_0 - \tilde{b}_1$.
\begin{equation} \label{eq:EQ013}
b_0 - \tilde{b}_1 = \mathcal{A}_{cc} - \Delta a_{cc} \hbar c \, \, \, ,
\end{equation}
where the isoscalar scattering length is to be thought of as untreated~\footnote{The untreated isoscalar scattering length $b_0$ also enters the strong shift $\epsilon_{1 s}$ in pionic deuterium. A combined analysis of the 
$\epsilon_{1 s}$ values in pionic hydrogen and deuterium, and of $\Gamma_{1 s}$ in pionic hydrogen enables a more accurate determination of the two scattering lengths (in comparison with the use of the information extracted 
only from pionic hydrogen), see Ref.~\cite{Schroeder2001,Sigg1995,Sigg1996a}.} (as the lack of the tilde over it indicates) and $\Delta a_{cc} = (-2.0 \pm 1.3) \cdot 10^{-3} \, m_c^{-1}$.

For the relation between $\Gamma_{1 s}$ of pionic hydrogen and the corresponding scattering length $\mathcal{A}_{c0}$, the authors used the expression:
\begin{align} \label{eq:EQ014}
\Gamma_{1 s}=4 q_0 \frac{\alpha^3 \mu^2}{(\hbar c)^2} \left( 1 + P^{-1} \right) \mathcal{A}_{c0}^2 \Big( & 1 + \varphi + 4 \alpha \left( 1 - \ln \alpha \right) \frac{\mu \mathcal{A}_{cc}}{\hbar c}\nonumber\\
& + 2 ( m_p + m_c - m_n - m_0 ) \frac{\mu b_0^2}{(\hbar c)^2} \Big) \, \, \, ,
\end{align}
where $m_n$ is the mass of the neutron and $\mathcal{A}_{c0} = \sqrt{2} \tilde{b}_1 + \Delta a_{c0} \hbar c$, with $\Delta a_{c0}=0.4(9) \cdot 10^{-3} \, m_c^{-1}$. Equations (\ref{eq:EQ013},\ref{eq:EQ014}) contain two 
unknowns: $b_0$ and $\tilde{b}_1$. The quantity $\tilde{b}_1$ may be evaluated by use of a simple recurrence relation; the convergence is fast. The quantity $b_0$ is subsequently obtained via Eq.~(\ref{eq:EQ013}). Evident 
from Refs.~\cite{Baru2011a,Baru2011b} is that the isospin-breaking effects have a larger impact on the isoscalar part of the $\pi N$ interaction at the $\pi N$ threshold. For this correction, the authors give the expression:
\begin{equation*}
\tilde{b}_0 = b_0 - \frac{m_p \hbar c}{m_p+m_c} \left( \frac{m_c^2 - m_0^2}{\pi F_\pi^2} c_1 - 2 \alpha f_1 \right) \, \, \, ,
\end{equation*}
where the values and uncertainties of the LECs $c_1$ and $f_1$, used in Refs.~\cite{Baru2011a,Baru2011b}, have already been given in Section 5.2.1.3. Comparison with Eq.~(\ref{eq:EQ012}) implies that the correction to $b_1$ 
reads as
\begin{equation} \label{eq:EQ014_5}
b_1 - \tilde{b}_1 = \frac{m_p \hbar c}{m_p+m_c} \frac{\alpha f_2}{2} \, \, \, ,
\end{equation}
and comes out equal to $-0.43(17) \cdot 10^{-3} \, m_c^{-1}$. Presumably, this correction is contained in $\Delta a_{cc}$ of Eq.~(\ref{eq:EQ013}). The corrected $\tilde{a}_{cc}$ may then be obtained as the difference 
$\tilde{b}_0-\tilde{b}_1$, whereas $\tilde{a}_{c0}=\sqrt{2} \tilde{b}_1$.

\subsubsection{\label{sec:FurtherRemarks}Comparison of results and last remarks on the corrections at the $\pi N$ threshold}

The important results of the application of the aforementioned correction schemes to $\epsilon_{1 s}$ and $\Gamma_{1 s}$, so that the two hadronic scattering lengths be extracted, are listed in Table \ref{tab:EMCorrections}. 
To enable a comparison with the results obtained with the models of Sections 5.2.1.1 and 5.2.1.2, and provide an impression of the largeness of the $\mathcal{O}(p^2)$ corrections in Ref.~\cite{Lyub2000}, a $\delta_\epsilon$ 
result was extracted after identifying $\mathcal{A}_{cc}$ with $\tilde{a}_{cc}$ or, equivalently, after ignoring the correction $\Delta \mathcal{A}_{cc}$ of Eq.~(\ref{eq:EQ012}). The difference between the corrections 
$\delta_\epsilon$ of Ref.~\cite{Lyub2000} and the corresponding value of Table \ref{tab:EMCorrections} is accounted for by the use of Eq.~(\ref{eq:EQ011}) herein, as opposed to Eq.~(\ref{eq:EQ010}) in Ref.~\cite{Lyub2000}.

\vspace{0.5cm}
\begin{table}[h!]
{\bf \caption{\label{tab:EMCorrections}}}The corrections to the measurements of the strong shift $\epsilon_{1 s}$ and the total decay width $\Gamma_{1 s}$ in pionic hydrogen. The input, common in all cases, comprises the 
average results for $\epsilon_{1s}$ and $\Gamma_{1s}$ of the two PSI experiments, with no uncertainties (statistical or systematic). All corrections are expressed in the form of $\delta_\epsilon$ and $\delta_\Gamma$, see 
Eqs.~(\ref{eq:EQ008},\ref{eq:EQ009}). A $\delta_\epsilon$ result was also obtained (last row in the upper part of the table) after identifying $\mathcal{A}_{cc}$ of Eq.~(\ref{eq:EQ011}) with $\tilde{a}_{cc}$.
\vspace{0.25cm}
\begin{center}
\begin{tabular}{|c|c|c|}
\hline
Source & $\delta_\epsilon$ ($10^{-2}$) & $\delta_\Gamma$ ($10^{-2}$)\\
\hline
\hline
\multicolumn{3}{|c|}{Methods aiming at the removal of the trivial EM effects}\\
\hline
\cite{Sigg1996b} & $-2.1(5)$ & $-1.3(5)$\\
\cite{Oades2007} & $0.67(67)$ & $-1.66(33)$\\
\cite{Ericson2004} & $-0.62(29)$ & $1.02(23)$\\
\hline
\cite{Lyub2000}, $\tilde{a}_{cc} \equiv \mathcal{A}_{cc}$ & $1.12$ &\\
\hline
\multicolumn{3}{|c|}{Methods aiming at the removal of EM effects,}\\
\multicolumn{3}{|c|}{as well as of effects due to $m_u \neq m_d$}\\
\hline
\cite{Lyub2000} & $-4.3 \pm 2.2$ & $-$\\
\cite{Gasser2002} & $-7.2 \pm 2.9$ & $-$\\
\cite{Zemp2004} & $-$ & $0.6(2)$\\
\cite{Baru2011a,Baru2011b} & $-7.2 \pm 2.6$ & $0.56(72)$\\
\hline
\end{tabular}
\end{center}
\vspace{0.5cm}
\end{table}

\begin{figure}
\begin{center}
\includegraphics [width=15.5cm] {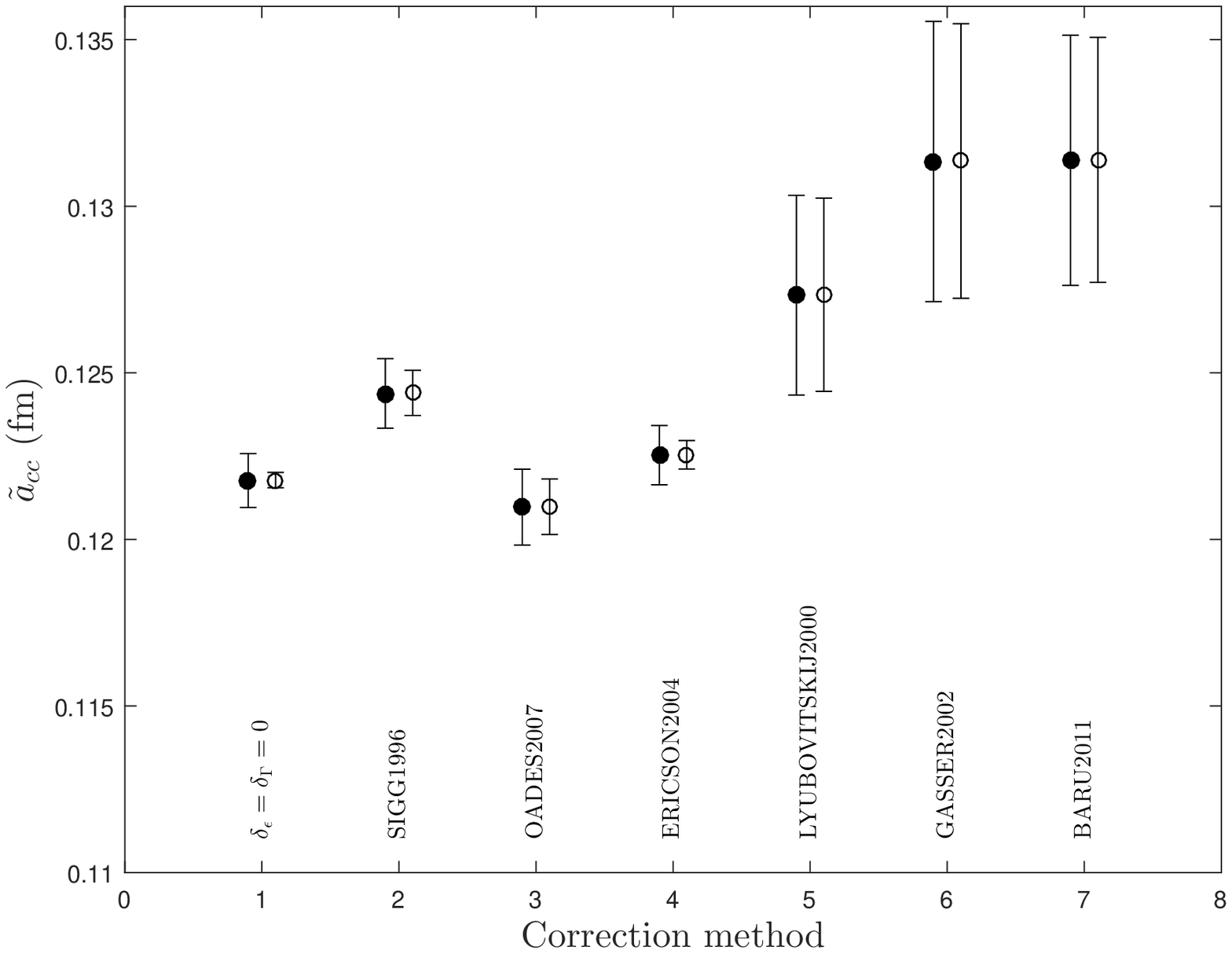}
\caption{\label{fig:Acc}The scattering length $\tilde{a}_{cc}$, obtained after the application of the correction methods of Sections 5.2.1.1-5.2.1.4 and 5.2.1.6 to the experimental results of the two PSI experiments. The 
last three correction schemes on the right take account of effects which go beyond the scope of the other three correction methods. Filled circles: Ref.~\cite{Schroeder2001}; open circles: Ref.~\cite{Hennebach2014}. To err 
on the side of caution, the statistical and systematic uncertainties of the experimental results were linearly combined for both experiments.}
\vspace{0.25cm}
\end{center}
\end{figure}

\begin{figure}
\begin{center}
\includegraphics [width=15.5cm] {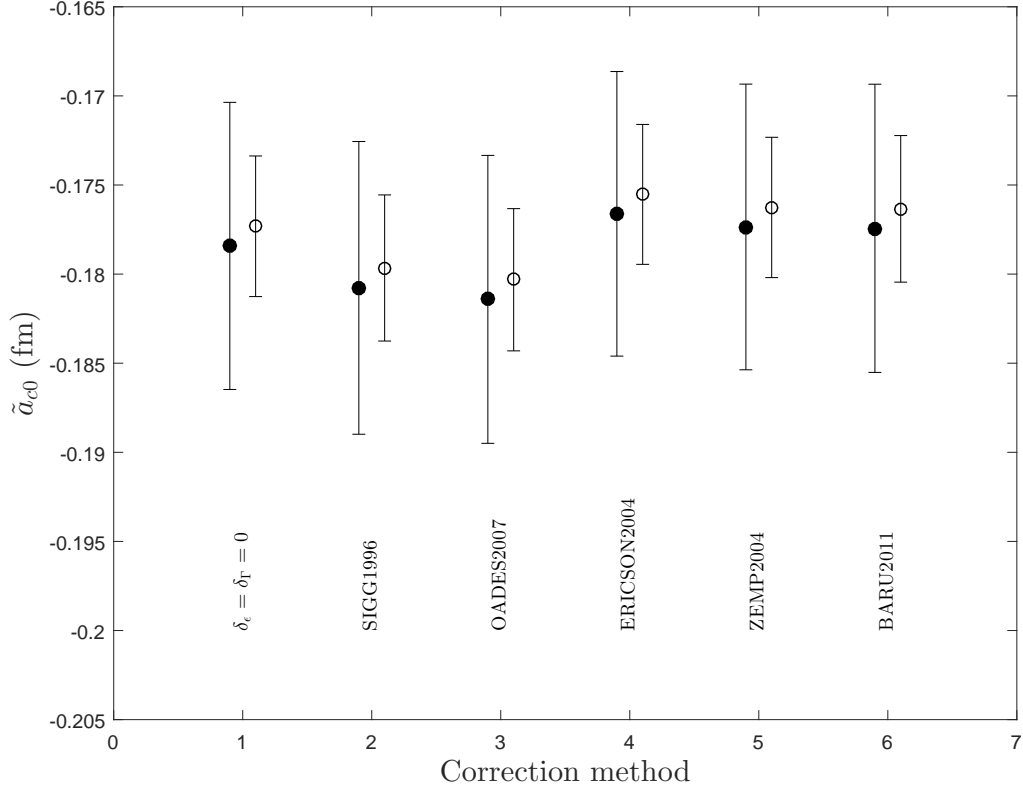}
\caption{\label{fig:Ac0}The scattering length $\tilde{a}_{c0}$, obtained after the application of the correction methods of Sections 5.2.1.1, 5.2.1.2, 5.2.1.5, and 5.2.1.6 to the experimental results of the two PSI 
experiments. The last two correction schemes on the right take account of effects which go beyond the scope of the other three correction methods. Filled circles: Ref.~\cite{Schroeder2001}; open circles: Ref.~\cite{Hirtl2021}. 
In both cases, the statistical and systematic uncertainties of the experimental results were linearly combined.}
\vspace{0.25cm}
\end{center}
\end{figure}

\begin{figure}
\begin{center}
\includegraphics [width=15.5cm] {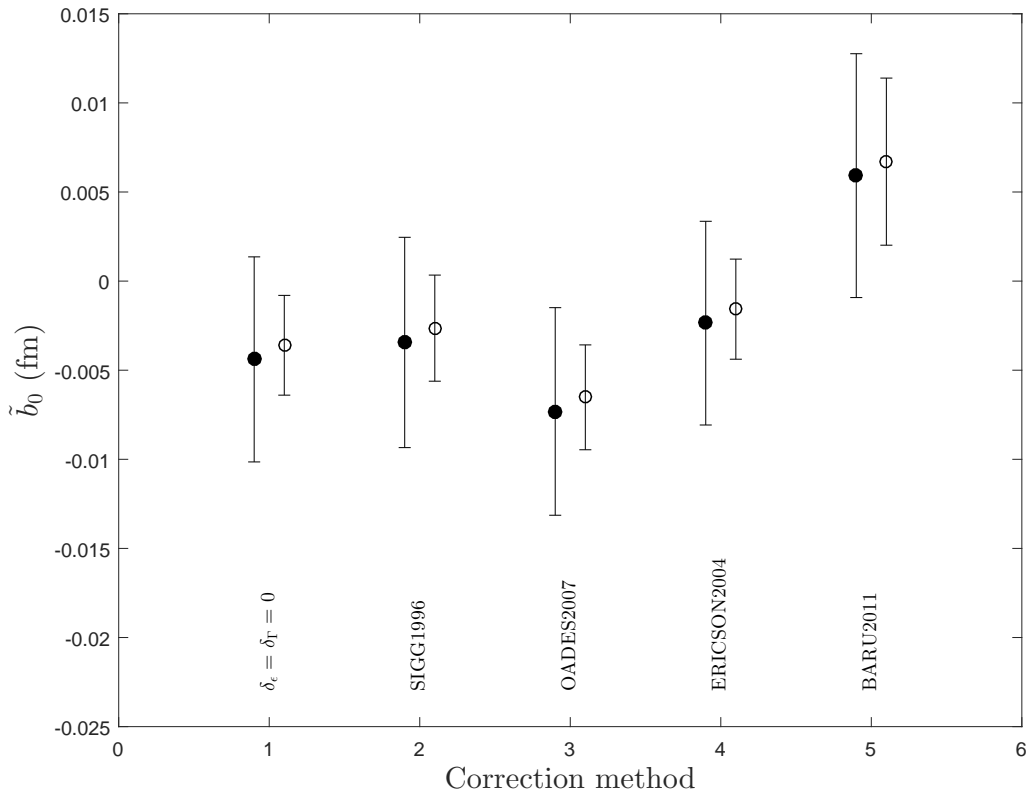}
\caption{\label{fig:b0}The isoscalar scattering length $\tilde{b}_0 \equiv \tilde{a}_{cc} + \tilde{a}_{c0}/\sqrt{2}$, obtained after the application of the correction methods of Sections 5.2.1.1, 5.2.1.2, and 5.2.1.6 to the 
experimental results of the two PSI experiments. The last correction scheme on the right take account of effects which go beyond the scope of the other correction methods. Filled circles: Ref.~\cite{Schroeder2001}; open 
circles: Refs.~\cite{Hennebach2014,Hirtl2021}. In both cases, the statistical and systematic uncertainties of the experimental results were linearly combined.}
\vspace{0.25cm}
\end{center}
\end{figure}

\begin{figure}
\begin{center}
\includegraphics [width=15.5cm] {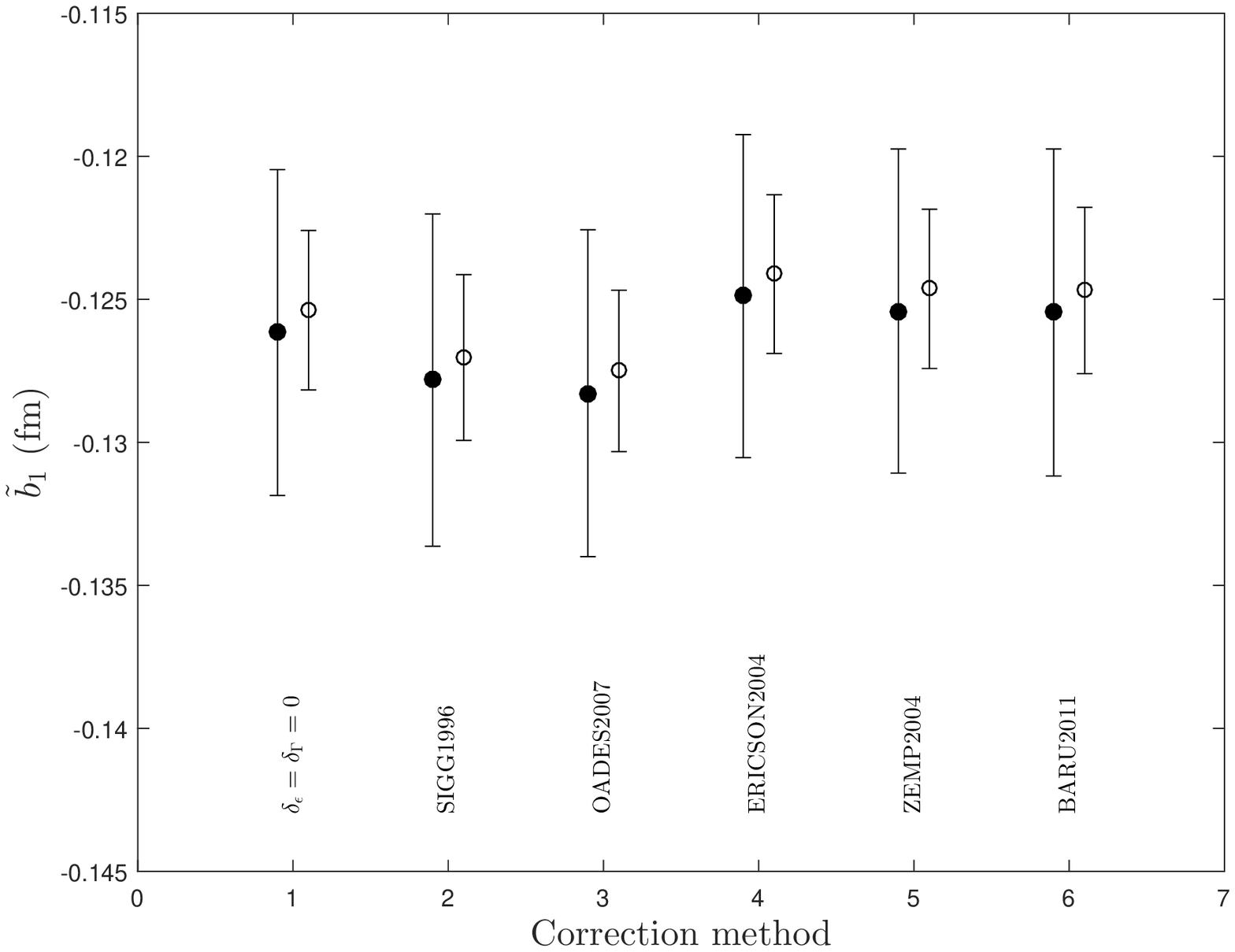}
\caption{\label{fig:b1}The isovector scattering length $\tilde{b}_1 \equiv \tilde{a}_{c0}/\sqrt{2}$, obtained after the application of the correction methods of Sections 5.2.1.1, 5.2.1.2, 5.2.1.5, and 5.2.1.6 to the 
experimental results of the two PSI experiments. The last two correction schemes on the right take account of effects which go beyond the scope of the other three correction methods. Filled circles: Ref.~\cite{Schroeder2001}; 
open circles: Refs.~\cite{Hennebach2014,Hirtl2021}. In both cases, the statistical and systematic uncertainties of the experimental results were linearly combined.}
\vspace{0.25cm}
\end{center}
\end{figure}

The values of the two scattering lengths $\tilde{a}_{cc}$ and $\tilde{a}_{c0}$, corrected for the EM effects using the methods of Sections 5.2.1.1 and 5.2.1.2, and for the EM effects and effects of hadronic origin of Sections 
5.2.1.3-5.2.1.6 are given in Figs.~\ref{fig:Acc} and \ref{fig:Ac0}, respectively. Shown in Figs.~\ref{fig:b0} and \ref{fig:b1} are the corresponding results for the isoscalar $\tilde{b}_0$ and the isovector $\tilde{b}_1$ 
scattering lengths, respectively. Inspection of these plots provides an impression as to the uncertainties in the original experimental data and the ones which the various correction schemes introduce; the uncertainties 
of the quantities $\alpha$ and $\mu$ in the two Deser formulae of Eqs.~(\ref{eq:EQ006},\ref{eq:EQ007}) are tiny, whereas the relative uncertainty of $P$ in the second of these equations is close to $6$ per-mille. In spite 
of the large uncertainty, noticeable is the change in the isoscalar part of the $\pi N$ interaction when applying to the pionic-hydrogen measurements the results of the method of Section 5.2.1.6.

Inspection of Table \ref{tab:EMCorrections} leads to the following conclusions.
\begin{itemize}
\item A consistent picture for the corrections $\delta_\epsilon$ and $\delta_\Gamma$ does not emerge from the upper part of this table (Methods aiming at the removal of the trivial EM effects).
\item One notices that the uncertainty of the $\delta_\epsilon$ correction of Ref.~\cite{Ericson2004} is sizeably smaller than those obtained with all other methods; this is due to the fact that only their gauge term is 
accompanied by an appreciable uncertainty. To obtain an estimate for that contribution, Ref.~\cite{Ericson2004} had imported information from the Karlsruhe programme of the 1980s \cite{Hoehler1983}, which is - to a great 
extent - outdated by today's standards; current information suggests that the values of the two range parameters, entering the contributions of the gauge term to $\delta_\epsilon$ and $\delta_\Gamma$, are (first) different 
and (second) better known \cite{Matsinos2022b}, in comparison with the values used in Ref.~\cite{Ericson2004}. As a result, the updated $\delta_\epsilon$ and $\delta_\Gamma$ values come out different now and are (even) 
better known; in my judgment, the uncertainties of Ref.~\cite{Ericson2004}, as well as those of the updated corrections \cite{Matsinos2022b}, are model-dependent and overly optimistic.
\item One may argue that the genuine three-channel calculation of Ref.~\cite{Oades2007} constitutes an improvement over the approach of Ref.~\cite{Sigg1996b}, and thus proceed to compare the $\delta_\epsilon$ and $\delta_\Gamma$ 
results of Ref.~\cite{Oades2007} with those obtained with the only other approach which does not deploy $\chi$PT, namely Ref.~\cite{Ericson2004}. Obviously, there is no matching; the signs are opposite in both corrections 
$\delta_\epsilon$ and $\delta_\Gamma$. Moreover, the discrepancy between the two corrections $\delta_\Gamma$ is disturbing.
\item The correction $\delta_\epsilon$ of Ref.~\cite{Oades2007} appears to be compatible with the result obtained from the upgraded form of the Deser formula for $\epsilon_{1 s}$ (see Eq.~(\ref{eq:EQ011})), whereas the 
corresponding result of Ref.~\cite{Ericson2004} is not. Nevertheless, it is not clear that such a comparison is meaningful. Part of the EM corrections of Refs.~\cite{Oades2007,Ericson2004} are contained in the upgraded form 
of the Deser formula for $\epsilon_{1 s}$; another part is contained in the correction $\Delta \mathcal{A}_{cc}$; a third part is not contained in the correction $\Delta \mathcal{A}_{cc}$. Therefore, the compatibility 
between the correction $\delta_\epsilon$ of Ref.~\cite{Oades2007} with the result obtained from the upgraded form of the Deser formula for $\epsilon_{1 s}$ may be coincidental.
\item The correction $\delta_\Gamma$ extracted in Ref.~\cite{Ericson2004} is not incompatible with the two estimates obtained within the $\chi$PT framework in Refs.~\cite{Baru2011a,Baru2011b,Zemp2004}; the same goes for 
the updated result of Ref.~\cite{Matsinos2022b}, which (in fact) is in perfect agreement with the estimates of Refs.~\cite{Baru2011a,Baru2011b,Zemp2004}. It has been suggested that potential models have a tendency to yield 
negative corrections $\delta_\Gamma$. Given the outcome of Refs.~\cite{Sigg1996b,Oades2007}, there might be some truth in this supposition.
\item It is time I discussed the corrections $\delta_\epsilon$ obtained within the $\chi$PT framework. The corrections $\delta_\epsilon$ of Refs.~\cite{Lyub2000,Gasser2002,Baru2011a,Baru2011b} are large and, even worse, 
poorly known. The sizeable uncertainties are mostly attributable to the poor knowledge of the LEC $f_1$. The difference between Refs.~\cite{Lyub2000,Gasser2002} is that, in the former work, the additional isospin-breaking 
effects are treated at $\mathcal{O}(p^2)$; in Ref.~\cite{Gasser2002}, they are treated at $\mathcal{O}(p^3)$. If, as the result of the application of the correction $\Delta \mathcal{A}_{cc}$ of Eq.~(\ref{eq:EQ012}), 
$\delta_\epsilon$ changes by as much as $-5.4~\%$ (i.e., from $+1.1~\%$ to $-4.3~\%$) and the result of the correction at the next order brings another $-2.9~\%$, then I do wonder what the calculation at $\mathcal{O}(p^4)$ 
might bring. I see no signs of convergence.
\item The evaluation of the corrections $\delta_\epsilon$ within the $\chi$PT framework (Sections 5.2.1.3, 5.2.1.4, and 5.2.1.6) also involves the mass difference of the two light quarks. Apart from the effects, which 
in the language of Refs.~\cite{Sigg1996b,Oades2007} constitute `mass differences of the particles in the initial and final states', such procedures remove contributions, which should be categorised herein under Section 
\ref{sec:IsospinBreaking}. Therefore, it is not surprising that the corrections $\delta_\epsilon$ of Refs.~\cite{Lyub2000,Gasser2002,Baru2011a,Baru2011b} come out sizeably larger than those of 
Refs.~\cite{Sigg1996b,Oades2007,Ericson2004}. One remark is in order. The deduction of the EM contributions from the light-quark masses is not addressed in Refs.~\cite{Lyub2000,Gasser2002,Baru2011a,Baru2011b}.
\item The only positive conclusion from the inspection of Table \ref{tab:EMCorrections} is the overall agreement of Refs.~\cite{Ericson2004,Baru2011a,Baru2011b,Zemp2004} regarding the magnitude of the correction $\delta_\Gamma$: 
they all suggest that this correction does not exceed $\approx 1~\%$. This implies that the isovector part of the $\pi N$ scattering amplitude is better known (in comparison with the isoscalar part).
\end{itemize}

At this point, I cannot resist one comment. Between 1990 and 1995, I had heard at least four prominent theorists lamenting the lack of precise experimental information at the $\pi N$ threshold. The ETHZ-Neuch{\^a}tel-PSI 
Collaboration delivered $\epsilon_{1 s}$ with an uncertainty well below $1~\%$, whereas both statistical and systematic uncertainties, reported by the Pionic-Hydrogen Collaboration, were at the level of $0.1~\%$. Such 
accuracy is unprecedented in Pion Physics. After this precise information became available, the theorists discovered that no competitive, up-to-date EM-correction scheme had been developed~\footnote{In fact, the lack of 
such a scheme was the motivation for Sigg and collaborators to set out to examine the EM effects in pionic hydrogen in Ref.~\cite{Sigg1996b}.} to enable the extraction of the important (hadronic) information from the 
experimental results. If the best Theory can do is to provide corrections at the $\pi N$ threshold which are one order of magnitude less precise than the experimental results, then my opinion remains that Theory must find 
a way of `narrowing the gap'.

To summarise, the agreement of the results of the available schemes of removal of the EM effects at the $\pi N$ threshold is poor at present.
\begin{itemize}
\item The overall incompatibility of the results of the various schemes of removal of the trivial EM effects at the $\pi N$ threshold (upper part of Table \ref{tab:EMCorrections}), ought to be explained.
\item Regarding the corrections, which have been developed within the $\chi$PT framework, a twofold improvement would be welcomed: to assess the convergence, the isospin-breaking corrections ought to be obtained at order 
$\mathcal{O}(p^4)$; they must also become more precise (after proposing ways, which are within the bounds of possibility, to narrow down the uncertainty of the LEC $f_1$). In this context, it ought to be understood that the 
determination of accurate corrections at the $\pi N$ threshold is undoubtedly a good cause, but not the only one; the data analysis calls for the development of corresponding corrections in the scattering region.
\end{itemize}
In one phrase: a unified scheme for the development of reliable EM corrections, \emph{applicable at the $\pi N$ threshold and in the scattering region}, seems (to me) to be the only step forwards. Without doubt, we must 
first decide what effects should be removed from the experimental data (or from quantities emerging thereof).

\subsection{\label{sec:IsospinBreaking}Violation of the isospin invariance}

The discrepancies, which the analyses of the low-energy $\pi N$ data have established \cite{Matsinos2017a,Gibbs1995,Matsinos1997}, may be taken to suggest that the isospin invariance is broken in the $\pi N$ interaction at 
a level exceeding the $\chi$PT expectations \cite{Hoferichter2010}. I left this option for the end as, admittedly, it is the most compelling one in Physics terms. Although the conclusions of the studies 
\cite{Gibbs1995,Matsinos1997} had not been received with boundless enthusiasm, one is tempted to raise the question: `Why should the isospin invariance hold in the first place?' After all, the bare masses of the $u$ and $d$ 
quarks \emph{are} different. It seems, therefore, that the right question to ask is not whether the isospin invariance is broken in the $\pi N$ interaction, but at which level it is.

Piekarewicz has suggested \cite{Piekarewicz1995} that the isospin breaking in the $N N$ interaction can ``originate from: (i) isovector-isoscalar mixing in the meson propagator - such as the $\rho^0-\omega$ mixing; (ii) 
isospin-breaking in the nucleon wavefunction - through the neutron-proton mass difference; and (iii) isospin-breaking in the meson-nucleon and photon-nucleon vertices - as in electromagnetic scattering.'' Piekarewicz added 
that the aforementioned isospin-breaking mechanisms (ii) and (iii) ``also operate in the $\pi N$ system.'' A few remarks are in order. As Piekarewicz suggested, the $\rho^0-\omega$ mixing does not seem to play a role in 
$\pi N$ ES \cite{Matsinos2018}, yet another isovector-isoscalar mixing, the one involving the QM admixture of the $\pi^0$ and the $\eta$ meson, a state with quantum properties $I^G (J^{PC}) = 0^+ (0^{-+})$ and a rest mass 
of $547.862(17)$ MeV \cite{PDG2020}, has long been known as potentially affecting the $\pi^- p$ CX reaction \cite{Cutkosky1979}. In the language of the models based on hadronic exchanges, one would categorise the sources 
of isospin-breaking effects in the $\pi N$ interaction as follows (in order of increasing importance, in my opinion):
\begin{itemize}
\item mass differences in Feynman diagrams involving different members of the isospin multiplets;
\item splitting effects in the various coupling constants and, presumably, vertex factors; and
\item Feynman diagrams containing a $\pi^0 - \eta$ transition vertex (see Fig.~\ref{fig:IsospinBreakingEtaPi0}). Regarding the $\pi^0 - \eta$ mixing mechanism, it would be interesting to also have an estimate for the potential 
difference between the coupling constants $g_{\eta p p}$ and $g_{\eta n n}$ (though this might take some time).
\end{itemize}
Some authors call the effects of the first type `static' or (less frequently) `kinematical', whereas those of the second are usually described as `dynamical'. The effects due to the $\pi^0 - \eta$ transition should be 
categorised as `dynamical'.

Already in 1995, Piekarewicz \cite{Piekarewicz1995} put forward an explanation for the surprising result of Ref.~\cite{Gibbs1995}; using a non-relativistic constituent-quark model, he evaluated the changes in the $\pi N$ 
coupling constant due to the mass difference of the $u$ and $d$ quarks. Regarding the discrepancy $D = \Re \left[ f^{\rm extr}_{\rm CX} - f_{\rm CX} \right]$ for the $s$ wave, which had come out equal to $-0.012(3)$ fm in 
Ref.~\cite{Gibbs1995}, Piekarewicz obtained in his paper the relation:
\begin{equation} \label{eq:EQ016}
D = - \sqrt{2} \, \hbar c \, \frac{g^2_{\pi N N}}{4 \pi} \frac{g^\pi_0}{M} \left( 1+\frac{m_c}{M} \right)^{-1} \left( 1-\frac{m^2_c}{4 M^2} \right)^{-1} \, \, \, ,
\end{equation}
where $g_{\pi N N}$ stands for the isospin-conserving coupling constant (for which $g^2_{\pi N N}/(4 \pi) = 14.21$ was used in Ref.~\cite{Piekarewicz1995}) and $M$ is the nucleon mass (average of the proton and neutron 
masses); $g^\pi_0$ denotes the isospin-violating component in the $\pi^0 N N$ coupling constant, for which Piekarewicz obtained the relation
\begin{equation*}
g^\pi_0 = 0.3 \frac{\Delta m}{m} \approx 0.004 \, \, \, ,
\end{equation*}
where $m$ and $\Delta m$ denote the average of the \emph{constituent} masses of the two light quarks ($m=M/3$) and their difference, respectively. In his evaluation, he chose $\Delta m = 4.1$ MeV, which actually coincides 
with the central value appearing in a peculiar inequality $m_d - m_u > (4.1 \pm 0.3)$ MeV \cite{Lichtenberg1989}, and obtained $D \approx -0.0145$ fm, i.e., a value compatible with the result of Ref.~\cite{Gibbs1995}. Fixing 
$\Delta m$ to $4$ MeV from Ref.~\cite{Griffiths2008}, see Table 4.4 therein (p.~135), and using the average $f^2_c$ value of Ref.~\cite{Matsinos2019}, one extracts from Eq.~(\ref{eq:EQ016}) about the same discrepancy which 
Piekarewicz had obtained in 1995 ($D \approx -0.0138$ fm).

The splitting effects in the $\pi N$ coupling constant were also studied two years later \cite{Meissner1997}: the authors suggested that $g_{\pi^\pm p n}$ should be equal to the average of the two $g_{\pi^0 N N}$ values and 
provided an estimate for the splitting $(g_{\pi^0 p p} - g_{\pi^0 n n})/g_{\pi^\pm p n}$. A short review on the status of the determination of the various $\pi N$ coupling constants may be found in Ref.~\cite{Matsinos2019}. 
The result of Ref.~\cite{Meissner1997} implies that $f_p>f_n$ or, equivalently, that $f^2_p>f^2_0$; the odds in favour of the opposite inequality come out to be about $2:1$ from Ref.~\cite{Matsinos2019} and about $3.5:1$ 
from Ref.~\cite{Reinert2021}. In 2017, two studies \cite{Babenko2017,Navarro2017} also reported sizeable (however contradicting) splitting effects in the $\pi N$ coupling constant.

It is interesting to note that the starting point in both studies \cite{Piekarewicz1995,Meissner1997} is the difference between the masses of the two light quarks (Ref.~\cite{Piekarewicz1995} makes use of the constituent 
masses, Ref.~\cite{Meissner1997} of the bare ones). The authors then proceed to evaluate the splitting effect which that difference would imply for the $\pi N$ coupling constant. Provided that I interpret the quantity 
$g^\pi_0$ of Ref.~\cite{Piekarewicz1995} correctly, the resulting effects differ sizeably between the two studies: the former work came up with a splitting effect of the order of $0.4~\%$, whereas the latter one reported 
effects between $1.2$ and $3.7~\%$. The former work concluded that the reported discrepancy $D$ of Ref.~\cite{Gibbs1995} could be accounted for, whereas the latter study did not examine that issue. I find it surprising 
that Ref.~\cite{Piekarewicz1995} could account for the discrepancy reported in Ref.~\cite{Gibbs1995} on the basis of a small splitting effect in the $\pi N$ coupling constant and without needing to invoke the $\pi^0 - \eta$ 
mixing mechanism.

Let me finally provide a synopsis of the four (thus far) $\chi$PT attempts to identify the source of the discrepancy reported in Refs.~\cite{Gibbs1995,Matsinos1997}; in none of these four papers did the authors refer to the 
previous work of Ref.~\cite{Piekarewicz1995}, which sought an explanation for the same discrepancy.

The first attempt by HB$\chi$PT to predict the level of the violation of the isospin invariance in the $\pi N$ interaction at low energy, on the basis of the mass difference of the two light quarks and the dominant 
virtual-photon effects, is detailed in Ref.~\cite{Fettes1999}. Eight indicators $R_{1 \dots 8}$ were used therein, to establish the type and quantify the level of the violation of the isospin invariance. One of these 
quantities, the indicator $R_2$, see Eq.~(\ref{eq:EQ002_5}) of this work, quantifies the amount of the departure from the triangle identity of Eq.~(\ref{eq:EQ002}). The results of the fits, appearing in their Table 1, 
suggest that their $R_2$ prediction is small, at the level of about $+1~\%$, i.e., not only of smaller magnitude in comparison with the results of Refs.~\cite{Gibbs1995,Matsinos1997}, but also opposite in sign. The authors 
commented that additional effort is needed, to include in the evaluation all virtual-photon effects, and pointed ``towards the necessity of a fourth-order calculation.''

Reference \cite{Fettes2001a} provides an extension of Ref.~\cite{Fettes1999} in terms of energy. Their Figs.~2 suggest that the indicator $R_2$ for the $s$ wave comes out about $-2.5~\%$, nearly constant in the low-energy 
region. Sizeable effects are seen in the $p$ waves, see their Figs.~3-6. The effect in the $s$ wave, albeit of the same sign now, still remains about three to four times smaller than the effects reported in 
Refs.~\cite{Gibbs1995,Matsinos1997}.

Reference \cite{Fettes2001b}, the last work by HB$\chi$PT for several years to follow, included all electromagnetic effects to third order. Regarding the violation of the isospin invariance, the authors wrote: ``The precise 
description of the scattering process also allows us to address the question of isospin violation in the strong interaction. For the usually employed triangle relation, we find an isospin breaking effect of $-0.7~\%$ in the 
$s$ wave, whereas the $p$ waves show effects of $-1.5~\%$ and $-4$ to $-2.5~\%$, respectively, for pion laboratory momenta between $25$ and $100$ MeV.'' Evidently, there seems to be another change in their $R_2$ prediction 
for the $s$ wave, leading to a less negative value in comparison with the results obtained in Ref.~\cite{Fettes2001a} by the same authors, similar in magnitude (but of opposite sign) to their first result \cite{Fettes1999}.

To my knowledge, the last work, addressing the issue of the violation of the isospin invariance in the $\pi N$ interaction within the $\chi$PT framework, may be found in Ref.~\cite{Hoferichter2010}. Using CB$\chi$PT and 
including all effects due to the mass difference of the two light quarks, as well as to real and virtual photons, to third order in the chiral expansion, the authors find substantial differences to the results of all earlier 
studies \cite{Fettes1999,Fettes2001a,Fettes2001b}, in both the $s$ and $p$ waves, see Section 3.3 therein. If their Fig.~18 is correct, then their upgraded corrections bring the indicator $R_2$ (for the $s$ wave) further 
away from the results of Refs.~\cite{Matsinos2017a,Gibbs1995,Matsinos1997}. In my judgment, it is counter-intuitive that $R_2$ (for the $s$ wave) increases (in magnitude) with increasing energy. At present, I do not see a 
coherent picture, emerging from these four attempts.

The subject of the isospin breaking in the $\pi N$ interaction may be considered in the light of what has long been known for the $N N$ system. Let me next touch upon the ways by which the isospin breaking has been 
established in the $N N$ interaction (see Ref.~\cite{Miller1990} for details).
\begin{itemize}
\item First, the hadronic part of the low-energy $N N$ interaction is characterised by three scattering lengths, corresponding to the $^1S_0$ states $p p$, $n n$, and $n p$. If the charge independence (which is used in the 
$N N$ domain as a synonym for the isospin invariance) held, then these three scattering lengths would have been equal. In reality, after the removal of the EM effects, their values are \cite{Miller2006}:
\begin{equation*}
a_{p p}=-17.3(4) \,\, {\rm fm}, \, \, \, a_{n n}=-18.8(3) \,\, {\rm fm}, \, \, \, a_{n p}=-23.77(9) \,\, {\rm fm} \, \, \, .
\end{equation*}
(In Ref.~\cite{Miller2006}, these scattering lengths carry the superscript `N', indicating that they are nuclear ones, obtained after the EM corrections had been applied.) Evidently, these values violate charge independence 
and, to a lesser extent, charge symmetry, as
\begin{equation*}
\Delta a_{CD}=(a_{p p}+a_{n n})/2-a_{n p}=5.7(3) {\rm fm}
\end{equation*}
and
\begin{equation*}
\Delta a_{CSD}=a_{p p}-a_{n n}=1.5(5) {\rm fm}
\end{equation*}
are significantly non-zero. These values correspond to the violation of the charge independence in the low-energy $s$-wave part of the $N N$ scattering amplitude by about $27~\%$ and of charge symmetry by about $8~\%$.
\item Second, there is a discrepancy between the theoretical expectations and the experimental results regarding the binding-energy differences of mirror nuclei. The effect was established in the 1960s \cite{Okamoto1964,Nolen1969} 
and is known as the `Okamoto-Nolen-Schiffer anomaly'.
\item Third, measurements of the spin-dependent left-right asymmetries in ES of polarised neutrons off polarised protons in the late 1980s and the early 1990s established significant differences between the APs of the two 
nucleons \cite{Abegg1986,Abegg1989,Knutson1990,Knutson1991}.
\end{itemize}

The level of the charge-independence breaking in the $N N$ interaction exceeds by far the magnitude of the effects which were reported in Refs.~\cite{Matsinos2017a,Gibbs1995,Matsinos1997} for the low-energy $\pi N$ 
interaction. If the $\pi N$ interaction is viewed as the basis for the description of the $N N$ interaction (as the case is in meson-exchange models of the $N N$ interaction), it is a plausible expectation that part of the 
(large) isospin-breaking effects, observed in the $N N$ system, could originate from the $\pi N$ interaction. In the mid 1990s, Gardner and collaborators had examined the impact of the splitting in the meson-baryon coupling 
constants, arising from the mass difference of the two light quarks, on the observables in the $N N$ interaction \cite{Gardner1995,Gardner1996}, and found out that the aforementioned discrepancy between the APs of the two 
nucleons could be accounted for.

At present, the source of the isospin-breaking effect in the $\pi N$ interaction at low energy, which was reported over twenty-five years ago \cite{Gibbs1995,Matsinos1997}, remains unknown (or even ``mysterious'' 
\cite{Fettes2001a}). The attempts to attribute the effect to the difference between the bare masses of the two light quarks, which have been carried out within the $\chi$PT framework, have not brought fruit yet, in that 
none of their predictions comes anywhere close to the level of the reported effects \cite{Matsinos2017a,Gibbs1995,Matsinos1997}; of course, it may turn out in the end that the origin of the effect involves more `mundane' 
phenomena than the mass difference of the two light quarks. Although the constituent-quark model of Ref.~\cite{Piekarewicz1995} has demonstrated that the effect may be understood as originating from the difference between 
the constituent masses of the $u$ and $d$ quarks, it would be desirable to obtain a corroboration from another - perhaps less phenomenological - approach, so that the source of the discrepancy be unequivocally identified.

\section{\label{sec:Conclusions}Conclusions}

Over twenty-five years ago, two analyses \cite{Gibbs1995,Matsinos1997} of the pion-nucleon ($\pi N$) data at low energy (i.e., for pion laboratory kinetic energy $T \leq 100$ MeV) reported on the departure of the extracted 
$\pi N$ scattering amplitudes, corresponding to the three reactions which are experimentally accessible at low energy (namely the two elastic-scattering (ES) reactions $\pi^\pm p \to \pi^\pm p$ and the $\pi^- p$ 
charge-exchange (CX) reaction $\pi^- p \to \pi^0 n$), from the `triangle identity' of Eq.~(\ref{eq:EQ002}), a relation which is fulfilled, if the isospin invariance holds in the $\pi N$ interaction. The aforementioned 
studies had used different ways of modelling the hadronic part of the $\pi N$ interaction, different schemes of application of the electromagnetic (EM) corrections, different (though overlapping) databases (DBs), and 
different methodologies. In spite of these differences, the two studies agreed well on the level of the departure from the triangle identity, thus giving rise to what I call herein the `low-energy $\pi N$ enigma'. This 
result has been corroborated several times by subsequent phase-shift analyses of the low-energy $\pi N$ data; as newer measurements emerged and were included in the DB, the statistical significance of the effect improved 
over time \cite{Matsinos2017a}. Point No.~1: \emph{The analysis of the low-energy $\pi N$ data gives rise to a significant departure of the extracted $\pi N$ scattering amplitudes from the triangle identity of 
Eq.~(\ref{eq:EQ002}).} The existence of this effect can hardly be disputed; its explanation, however, is open to speculation.

The first question arises: can global fits to the entirety of the $\pi N$ data be performed? Section \ref{sec:JointFits} starts with a positive answer to this question. The question however, whether or not such fits are 
satisfactory, admits a negative answer. For the sake of example, examined herein was one such popular solution, the `current' (XP15) solution of the SAID website \cite{XP15}. When the Arndt-Roper formula \cite{Arndt1972} is 
used in the optimisation (see Eq.~(\ref{eq:EQ003})), the statistical expectation is that the datasets which are scaled `upwards' balance (on average) those which are scaled `downwards'. Furthermore, the energy dependence of 
the scale factors $z_j$ of Eq.~(\ref{eq:EQ004}) must not be significant. The fitted values of the scale factor $z$ in the XP15 solution (for $T \leq 100$ MeV) show no significant bias, see Fig.~\ref{fig:ALLXP15}: the linear 
fit results in an intercept which is compatible with $1$ and a slope value which is compatible with $0$, see Table \ref{tab:XP15Parameters}. Therefore, the output of their global fit passes the first test of compatibility 
with the statistical expectation (i.e., with an unbiased outcome of optimisations resting upon the Arndt-Roper formula \cite{Arndt1972}). However, the SAID $\pi N$ DB comprises three distinct parts, i.e., the sets of 
observations in the three low-energy $\pi N$ reactions. Had their global fit been truly satisfactory, a similar behaviour of the scale factors $z_j$ (in terms of their energy dependence), which is seen in Fig.~\ref{fig:ALLXP15}, 
should also have been observed in any arbitrary subset of their DB, complying with the basic principles of the Sampling Theory (adequate population, representative sampling). The fitted values of the scale factor $z$, 
relating to the description of the SAID low-energy $\pi N$ DBs with the XP15 solution, are shown (separately for the three reactions) in Figs.~\ref{fig:PIPELXP15}-\ref{fig:PIMCXXP15}. If the global fit were truly satisfactory, 
then the scale factors $z_j$ of Figs.~\ref{fig:PIPELXP15}-\ref{fig:PIMCXXP15} would have come out independent of the beam energy and would have been centred on $1$ (as the case was for the results of the analysis of all 
scale factors $z_j$, shown in Fig.~\ref{fig:ALLXP15}). However, the bulk of the data for $T \leq 100$ MeV (represented by the shaded bands in Figs.~\ref{fig:PIPELXP15}-\ref{fig:PIMCXXP15}) seem to be either underestimated 
by the XP15 solution (i.e., in case of the $\pi^- p$ CX reaction) or overestimated by it (i.e., in case of the two ES reactions, the effects for the $\pi^+ p$ reaction being more pronounced), see also Table 
\ref{tab:XP15Parameters}. One notices that the mismatches decrease with increasing beam energy, converging to $z=1$ in the vicinity of $T=100$ MeV. Such behaviour is in general agreement with the conclusions of 
Refs.~\cite{Matsinos2017a,Gibbs1995,Matsinos1997} for an energy-dependent isospin-breaking effect. From Figs.~\ref{fig:PIPELXP15}-\ref{fig:PIMCXXP15} and Table \ref{tab:XP15Parameters}, one may conclude that the XP15 
solution does not describe sufficiently well the bulk of the low-energy measurements. Nearly identical results in case of the former solution of the SAID group, the WI08 solution \cite{Arndt2006}, were reported in 
Ref.~\cite{Matsinos2017b}. Point No.~2: \emph{In global fits of isospin-invariant analyses using the Arndt-Roper formula \cite{Arndt1972}, the low-energy $\pi N$ enigma gives rise to systematic energy-dependent effects in 
the scale factors $z_j$ of Eq.~(\ref{eq:EQ004}), which (as expected) are different for the three low-energy $\pi N$ reactions.}

Further analysis of the low-energy $\pi N$ data demonstrates that the joint optimisation of the ES measurements yields unbiased results, in accordance with the statistical expectation, see Ref.~\cite{Matsinos2017a} and the 
works cited therein. Therefore, there is no indication at present of violation of the isospin invariance in the two ES reactions. The departure from the statistical expectation occurs when one attempts to also include in 
the fit the $\pi^- p$ CX data (thus pursuing a global fit of the data of all three reactions). For that reason, it seems that the low-energy $\pi N$ enigma is due to the inability to accommodate the measurements of the 
$\pi^- p$ CX reaction in a global fit.

The first (perhaps naive) attempt to provide an explanation for the low-energy $\pi N$ enigma involves an obvious effect, namely the systematic inaccuracy of the absolute normalisation of the low-energy $\pi N$ datasets. 
However, this issue cannot be addressed by analysts, but should rather be resolved by the experimental groups which had been responsible for the measurements. At present, such a re-examination, even of a small fraction of 
the available datasets (e.g., of those datasets which look suspicious either because of their absolute normalisation or because of the smallness of their published normalisation uncertainty), seems to be a remote prospect. 
Point No.~3: \emph{It is unlikely that the reassessment of the correctness of the absolute normalisation of the datasets of the low-energy $\pi N$ DB will (or, in all probability, even could) take place. Given that this 
possibility seems to be barred, I see no other option (for the time being) but to accept the correctness of the absolute normalisation of the bulk of the experimental data at low energy.} On the other hand, there is strong 
indication that the normalisation uncertainties had been seriously underestimated \emph{on average} in the $\pi N$ experiments at low energy, in particular in the $\pi^+ p$ experiments, see Table \ref{tab:XP15Parameters}.

The second attempt to provide an explanation for the low-energy $\pi N$ enigma involves significant residual contributions in the results of the various available EM-correction schemes in the scattering region 
\cite{Gibbs1998,Tromborg1976,Tromborg1977,Tromborg1978,Gashi2001a,Gashi2001b}. Such schemes aim at the removal of the EM effects from the $\pi N$ phase shifts and partial-wave amplitudes, and lead to the extraction of the 
(important) hadronic quantities, e.g., of the parameters of the hadronic potentials in Ref.~\cite{Gibbs1995}, of the parameters of the hadronic model of Ref.~\cite{Matsinos1997}, etc. The results of the aforementioned 
EM-correction schemes were compared in Ref.~\cite{Gibbs2005}; two conclusions were drawn therein:
\begin{itemize}
\item the differences among the results of the tested schemes are small and
\item the low-energy $\pi N$ enigma seems to be several times more pronounced than the impact which the EM corrections have on the analyses at low energy.
\end{itemize}
Consequently, one could draw the tentative conclusion that an explanation for the low-energy $\pi N$ enigma in terms of significant contributions missing from the established schemes of removal of the EM effects from the 
scattering data \cite{Gibbs1998,Tromborg1976,Tromborg1977,Tromborg1978,Gashi2001a,Gashi2001b} does not seem to be credible.

Can the situation at the $\pi N$ threshold provide some insight into the matter? This issue is addressed in detail in Section \ref{sec:ResidualEMAtThreshold}. Very precise measurements of the strong-interaction shift 
$\epsilon_{1 s}$ \cite{Schroeder2001,Hennebach2014} and less-precise ones of the total decay width $\Gamma_{1s}$ \cite{Schroeder2001,Hirtl2021} of the ground state in pionic hydrogen were obtained at the Paul Scherrer 
Institut (PSI) between the mid 1990s and the early 2000. To extract the $\pi N$ scattering lengths, corresponding to the $\pi^- p$ ES and CX reactions, one must remove the effects of EM origin. To this end, three schemes 
were developed between 1996 and 2007: two potential-model approaches \cite{Sigg1996b,Oades2007} (the latter study may be thought of as an extension of the former) and one method relying on the use of Coulomb wavefunctions, 
with a short-range strong interaction and extended charge distributions \cite{Ericson2004}. Although these schemes aim at the removal of the same effects (i.e., of the so-called trivial EM effects), their results mismatch.

In addition to the removal of the EM effects, corrections have surfaced, which also attempt to remove from the pionic-hydrogen measurements effects of hadronic origin (i.e., effects emanating from the mass difference of 
the two light quarks); such studies have been carried out within the framework of the Chiral-Perturbation Theory ($\chi$PT) \cite{Lyub2000,Gasser2002,Baru2011a,Baru2011b,Zemp2004}. The corrections to the $\pi^- p$ ES length 
in Refs.~\cite{Lyub2000,Gasser2002,Baru2011a,Baru2011b} are larger (in magnitude) than those obtained from potential models \cite{Sigg1996b,Oades2007} or from the model of Ericson and collaborators \cite{Ericson2004}, and 
(furthermore) they are accompanied by sizeably larger uncertainties, see Table \ref{tab:EMCorrections}. Evidently, there is no consensus at present even on what effects should be removed from the accurate PSI measurements.

In view of the incompatibility of the results of the various schemes of removal of the trivial EM effects at the $\pi N$ threshold (upper part of Table \ref{tab:EMCorrections}), it cannot be excluded that the established 
schemes of application of the EM corrections to the scattering data fail to remove a sizeable part of effects of EM origin. This is why Section \ref{sec:ResidualEMAtThreshold} concludes with a recommendation: to establish 
a unified scheme for the development of reliable EM corrections, applicable at the $\pi N$ threshold and in the scattering region. Point No.~4: \emph{Are significant contributions missing from the established schemes of 
removal of the EM effects from the scattering data \cite{Gibbs1998,Tromborg1976,Tromborg1977,Tromborg1978,Gashi2001a,Gashi2001b}?}

The third attempt to provide an explanation for the low-energy $\pi N$ enigma is via the violation of the isospin invariance in the $\pi N$ interaction at low energy. The question of the isospin breaking in the $\pi N$ 
interaction may be considered in the light of the large breaking effects, which have long been established in the $N N$ system \cite{Miller2006}. The argument resting upon the inevitable question `if the $N N$ system is 
affected, then why should the $\pi N$ system not be?' might be a convincing one, but even more convincing would have been the elimination of the possibility of a systematic bias in the absolute normalisation of the low-energy 
$\pi N$ datasets and the application of reliable EM corrections to the entirety of the $\pi N$ data, from the $\pi N$ threshold to the few-GeV region.

I will finalise this study with a few comments on issues which have troubled me over the years, listed below in order of decreasing importance.

The most troubling question is why the analysis of the DCSs of the CHAOS collaboration \cite{Denz2006} is not possible within the context of the ETH $\pi N$ project \cite{Matsinos2013,Matsinos2015}. As the beam energy in 
the CHAOS DCSs is low, an investigation of the description of these data within the $\chi$PT framework (e.g., using the method of Ref.~\cite{Alarcon2013}) or following the method of Ref.~\cite{Hoferichter2016} should be 
possible. Given the availability of the data for over one and a half decades, I find it surprising that this issue has not been pursued yet.

The process of the identification of the outliers in the DB has been perfected in the analyses carried out within the ETH $\pi N$ project \cite{Matsinos2017a}. To reduce the model dependence of the process, the polynomial 
parameterisation of the $s$- and $p$-wave $K$-matrix elements (i.e., the ETH parameterisation in the language of Ref.~\cite{Matsinos2022a}) is exclusively used in that task. On the whole, I am fond of the use of robust 
methods in fits to data containing outliers, in particular of methods which apply soft weights to all datapoints, different at each step of the optimisation, depending on their distance (at that step) to the bulk of the 
fitted values. Such a dynamical optimisation process enables one to retain the initial input DB, allowing all datapoints to be present at all steps of the optimisation, with suitable weights regulating their contribution to 
the minimisation function. To come up with a reliable robust version of the Arndt-Roper formula has been a task which I have not been able to accomplish yet.

As the $K$-matrix approach is followed by both methods used in the modelling of the hadronic part of the $\pi N$ interaction within the ETH project, the unitarity of the $\pi N$ partial-wave amplitudes is fulfilled. Although 
the ETH model is occasionally categorised as a `tree-level model', the association of the $K$-matrix elements and the partial-wave amplitudes via the unitarisation scheme does take account of higher-order effects; one way of 
understanding this is by considering the low-energy expansion of the $\pi N$ partial-wave amplitudes: $\mathscr{F}_{IJ} = K_{IJ} \left( 1 - i q K_{IJ} \right)^{-1} = K_{IJ} \left( 1 + i q K_{IJ} - \mathcal{O}(q^2) \right)$. 
The first term within the brackets would correspond to the tree-level approximation, the second to Feynman diagrams with one loop, and so on. The continuation of the unitarisation scheme into the unphysical region has been 
one question to which I have found no satisfactory solution yet.

At present, only the $s$- and $p$-wave $K$-matrix elements of the ETH model are used in the analyses; the (small) $d$- and $f$-wave contributions are imported from the SAID analyses (their current solution), which (given 
that they make use of high-energy data) are expected to be reliable. The contribution of the Feynman diagrams of the ETH model to the $d$- and $f$-wave amplitudes are known since 1993, yet the contributions from the six 
well-established $d$ and $f$ higher baryon resonances with masses below $2$ GeV and known branching fractions to $\pi N$ decay modes introduce additional parameters, which neither can be treated as free in the optimisation 
nor can they be fixed from external sources. These states are:
\begin{itemize}
\item $N(1520)~(3/2)^-$ ($D_{13}$),
\item $N(1675)~(5/2)^-$ ($D_{15}$),
\item $N(1680)~(5/2)^+$ ($F_{15}$),
\item $\Delta(1700)~(3/2)^-$ ($D_{33}$),
\item $\Delta(1905)~(5/2)^+$ ($F_{35}$), and
\item $\Delta(1950)~(7/2)^+$ ($F_{37}$).
\end{itemize}
Although the issue is of no practical significance, the appropriate inclusion of these effects would reduce the amount of information which needs to be imported into the analyses, featuring the ETH model, from external 
sources.

\begin{ack}
This research programme has been shaped to its current form after the long-term interaction with B.L.~Birbrair (deceased), A.~Gashi, P.F.A.~Goudsmit, A.B.~Gridnev, H.J.~Leisi, G.C.~Oades (deceased), G.~Rasche, and W.S.~Woolcock 
(deceased). I am deeply grateful to them for their contributions.

I am indebted to G.~Rasche for our numerous discussions on issues relating to this work, to many other subjects in Pion/Hadronic Physics, as well as to broader matters regarding the Philosophy of Science.

%
The Feynman diagrams in this paper were created with the software package JaxoDraw \cite{Binosi2004,Binosi2009}, available from jaxodraw.sourceforge.net. The remaining figures were created with MATLAB$^{\textregistered}$~(The 
MathWorks, Inc., Natick, Massachusetts, United States).
\end{ack}

\end{document}